\documentclass[fleqn,usenatbib]{mnras}

\usepackage{newtxtext,newtxmath}

\usepackage[T1]{fontenc}

\DeclareRobustCommand{\VAN}[3]{#2}
\let\VANthebibliography\thebibliography
\def\thebibliography{\DeclareRobustCommand{\VAN}[3]{##3}\VANthebibliography}

\usepackage{graphicx}	
\usepackage{amsmath}	
\usepackage{xcolor}
\usepackage{threeparttable}
\usepackage{url}
\usepackage{hyperref}

\newcommand{\ang}{\text{\AA}}
\newcommand{\msun}{{\rm M}_\odot}
\newcommand{\sbunit}{\rm erg \, s^{-1}\, cm^{-2}\,arcsec^{-2}}

\newcommand{\vect}[1]{\boldsymbol{#1}}
\newcommand{\oii}{[\ion{O}{II}]\,$\lambda\lambda\,3727,3729$ }
\newcommand{\oiii}{[\ion{O}{III}]\,$\lambda\,5008$ }
\newcommand{\lya}{Ly$\alpha$ }

\title[Turbulence in QSO nebulae]{Empirical constraints on the turbulence in QSO host nebulae from velocity structure function measurements}

\author[M. C. Chen et al.]{Mandy C. Chen$^{1}$\thanks{E-mail: mandychen@astro.uchicago.edu}, Hsiao-Wen Chen$^{1}$, Michael Rauch$^{2}$, Zhijie Qu$^1$,  Sean D. Johnson$^3$,
\newauthor
Jennifer I-Hsiu Li$^3$, Joop Schaye$^4$, Gwen C. Rudie$^2$, Fakhri S. Zahedy$^2$, Erin Boettcher$^{5,6,7}$, 
\newauthor
Kathy L. Cooksey$^8$, and Sebastiano Cantalupo$^9$
\\
$^{1}$Department of Astronomy and Astrophysics, The University of Chicago, Chicago, IL 60637, USA\\
$^{2}$The Observatories of the Carnegie Institution for Science, 813 Santa Barbara Street, Pasadena, CA 91101, USA\\
$^{3}$Department of Astronomy, University of Michigan, Ann Arbor, MI 48109, USA\\
$^{4}$Leiden Observatory, Leiden University, PO Box 9513, NL-2300 RA Leiden, the Netherlands\\
$^{5}$Department of Astronomy, University of Maryland, College Park, MD 20742, USA\\
$^{6}$X-ray Astrophysics Laboratory, NASA/GSFC, Greenbelt, MD 20771, USA\\
$^{7}$Center for Research and Exploration in Space Science and Technology, NASA/GSFC, Greenbelt, MD 20771, USA\\
$^{8}$Department of Physics and Astronomy, University of Hawai’i at Hilo, Hilo, HI 96720, USA\\
$^{9}$Department of Physics, University of Milan Bicocca, Piazza della Scienza 3, I-20126 Milano, Italy
}

\date{Accepted XXX. Received YYY; in original form ZZZ}

\pubyear{2022}

\begin{document}
\label{firstpage}
\pagerange{\pageref{firstpage}--\pageref{lastpage}}
\maketitle

\begin{abstract}
We present the first empirical constraints on the turbulent velocity field of the diffuse circumgalactic medium around four luminous QSOs at $z\!\approx\!0.5$--1.1. Spatially extended nebulae of $\approx\!50$--100 physical kpc in diameter centered on the QSOs are revealed in \oii and/or \oiii emission lines in integral field spectroscopic observations obtained using MUSE on the VLT.  We measure the second- and third-order velocity structure functions (VSFs) over a range of scales, from $\lesssim\!5$ kpc to $\approx\!20$--50 kpc, to quantify the turbulent energy transfer between different scales in these nebulae.  While no constraints on the energy injection and dissipation scales can be obtained from the current data, we show that robust constraints on the power-law slope of the VSFs can be determined after accounting for the effects of atmospheric seeing, spatial smoothing, and large-scale bulk flows.  Out of the four QSO nebulae studied, one exhibits VSFs
in spectacular agreement with the Kolmogorov law, expected for isotropic, homogeneous, and incompressible turbulent flows. The other three fields exhibit a shallower decline in the VSFs from large to small scales. However, with 
a limited dynamic range in the spatial scales in seeing-limited data, no constraints can be obtained for the VSF slopes of these three nebulae. For the QSO nebula consistent with the Kolmogorov law, we determine a turbulence energy cascade rate of $\approx\!0.2$ cm$^{2}$ s$^{-3}$.  We discuss the implication of the observed VSFs in the context of QSO feeding and feedback in the circumgalactic medium.

\end{abstract}




\section{Introduction}
The tenuous gas residing in the circumgalactic medium (CGM) contains a critical record of the past and ongoing interactions between galaxies and their surrounding environment.  Characterizing the detailed physical properties of the CGM is an important step in improving current galaxy evolution models. Over the past three decades, absorption spectroscopy using predominantly QSO sightlines has yielded sensitive constraints on various properties of the CGM, and provided us with an increasingly intricate picture of the gaseous halo ecosystem \citep[see e.g.][and references therein]{Chen2017, Tumlinson2017, Rudie2019}. Observations have shown that the CGM contains multiphase gas spanning a wide range in density, temperature, ionization state, and metallicity \citep[e.g.,][]{Savage2005,Zahedy2019,Zahedy2021,Cooper2021}. Numerical simulations have also shown that different dynamical processes, such as gas infall, outflow, and tidal interactions, can also happen in the CGM to drive and regulate galaxy growth over cosmic time \citep[e.g.,][]{vandeVoort2017,Angles-Alcazar2017,Mitchell2022}. 

However, the lack of spatial information from the ``pencil-beam" probe of absorption spectroscopy has hindered our ability to robustly characterize the thermodynamic state of the gas. While the Doppler width of absorption profiles exceeding the value of thermal broadening may provide evidence for the presence of non-thermal pressure support in the CGM \citep[e.g.][]{Rauch1996,Rudie2019}, interpretations of the physical origin of the non-thermal motions remain ambiguous because both large-scale coherent flows and turbulent motions contribute to the observed line broadening.  Similar ambiguities exist in kinematic studies of emission signals obtained through long-slit or single-aperture spectroscopy. 

Diffuse, ionized plasmas such as the CGM are expected to be turbulent, because of the expected high Reynolds number \cite[see][for a recent review]{Burkhart2021}.  The presence of turbulence in the diffuse halo gas and the degree of such turbulence have profound implications for the thermal and dynamic properties of the CGM. Turbulent energy can be a significant source of heating to offset cooling in the hot halo through non-linear interactions between large and small eddies \citep[e.g.,][]{MacNamara2007,Zhuravleva2014}. In addition, turbulence produces density fluctuations, triggering and facilitating multiphase condensation in the hot halo \citep[e.g.,][]{Gaspari2018,Fielding2020,Gronke2022}. Turbulent mixing also provides an efficient transport mechanism for metals from star-forming regions to the CGM/IGM, and can facilitate the mixing of metals within the CGM \citep[e.g.][]{Pan2010}. Given these vital scientific implications, it is of great interest to obtain direct empirical constraints on turbulence in the CGM. 

\begin{figure*}
	\includegraphics[width=0.95\textwidth]{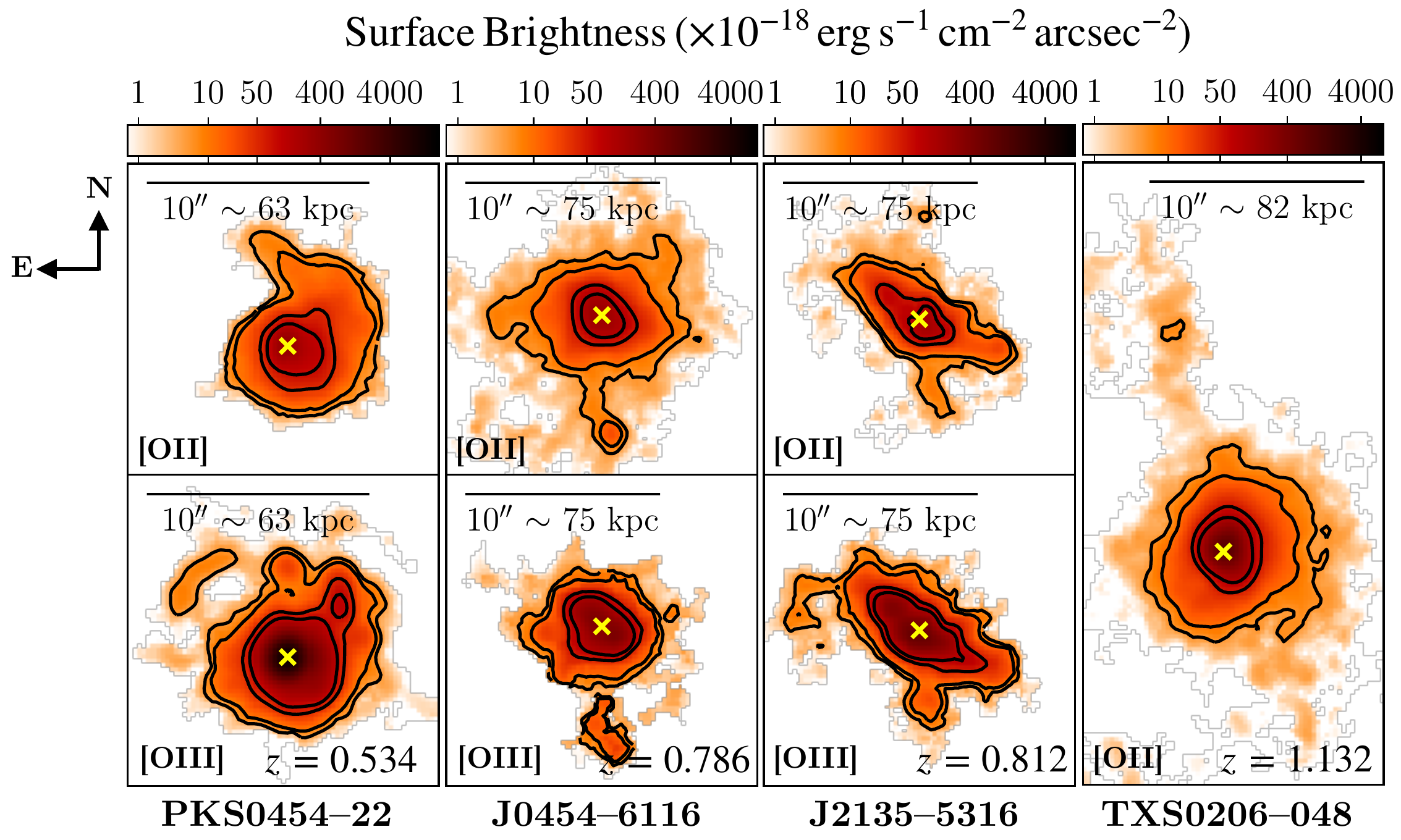}
    \caption{Continuum- and QSO-subtracted narrow-band images of the [\ion{O}{II}] and [\ion{O}{III}] emission from the four fields studied in this paper, based on the MUSE-WFM observations. The fields are shown in the order of increasing redshift from left to right.  For TXS0206$-$048, the [\ion{O}{III}] line is redshifted out of the MUSE wavelength coverage and is therefore not shown here.  Contours are at surface brightness levels of $[5, 10, 50, 100]\times 10^{-18}\,{\rm erg\,s^{-1}\, cm^{-2}\,arcsec^{-2}}$. The yellow cross in each panel marks the quasar position. }
    \label{fig:nb}
\end{figure*}

In this context, the recent advent of high-throughput, wide-field integral field spectrographs (IFSs) such as the Multi-Unit Spectroscopic Explorer \citep[MUSE;][]{Bacon2010} on the Very Large Telescope (VLT) 
has 
transformed CGM investigations by providing two-dimensional contiguous maps of large-scale line-emitting signals 
with unprecedented sensitivities and efficiency. Compared with absorption spectroscopy, the spatial information provided by these IFS data reveals new insights into the detailed physical processes of these low-density regions. In particular, spatially-resolved kinematic properties now enable two-point statistical measurements of the velocity field, providing an exciting opportunity to probe turbulence beyond 
a single sightline/aperture approach.  

One of the standard two-point probes is the velocity structure function (VSF), defined as
\begin{equation}
    S_p(r)=\langle | \vect{v}(\vect{x}) - \vect{v}(\vect{x}+\vect{r}) |^p \rangle,
	\label{eq:vsf}
\end{equation}
where $\vect{x}$ and $\vect{r}$ represent respectively the spatial location of a starting point and the distance between the starting point and a second location for calculating the pair velocity difference \citep[e.g.,][]{Frisch1995}. 
Different variants of the VSFs record the mean pair velocity difference to the power of $p$ averaged over all available pair configurations for a given separation $r\equiv|\vect{r}|$.
There have been extensive efforts, both in observations and numerical simulations, in using VSFs to probe the thermodynamic state of the interstellar medium (ISM) in local H\,{\small II} regions and in molecular clouds \citep[e.g.][]{Wen1993,Ossenkopf2002,Federrath2013,Padoan2016,Arthur2016,AnorveZeferino2019,Chira2019,Melnick2021,HuYue2022MNRAS}. Recently, VSFs have also been measured for Milky Way stars using GAIA data \citep[][]{Ha2021, Ha2022}. These studies in the local Universe have shown that not only is turbulence ubiquitous in the ISM, but it also plays a critical role in shaping the star-formation processes inside a galaxy \citep[see, e.g.][for a review]{Burkhart2021}. 

For the CGM, measuring VSFs becomes more challenging because of faint emission signals and because of a lack of two-dimensional velocity maps with sufficiently fine spatial sampling, particularly for sources beyond the local Universe where cosmological surface brightness dimming further weakens the signal strength. \cite{Rauch2001} attempted the first second-order VSF measurements in the low-density circum- and intergalactic gas at redshift $z\approx 2-3$, using C\,{\small IV} absorbers identified along multiply-lensed QSO sightlines. The VSFs in \cite{Rauch2001} were found to be consistent with expectations from the Kolmogorov turbulence \citep[][also see the discussion below in \S\,\ref{sec:VSF_bkg}]{Kolmogorov1941}, but the uncertainties were large  and 
the spatial sampling was sparse. Recently, \cite{Li2020} measured the first-order VSF using H$\alpha$ filaments detected in IFS data near the centers of nearby cool core clusters.  These authors identified a bump in the VSFs at 20-30 kpc, which they attributed to energy injections by rising bubbles powered by  
the supermassive black holes at the centers of  
these galaxy clusters.  Studies such as these demonstrate that measuring the VSF provides a promising tracer of energy coupling and cascades from the source at the galactic center to the diffuse gas reservoir on 10--30\,kpc scales.

Motivated by \cite{Li2020}, we have carried out a detailed analysis of the velocity field observed 
in four QSO-host nebulae.  These nebulae are revealed by extended emission (up to a scale of $\sim 100$ physical kpc in diameter) in \oii and/or \oiii lines (see Figure \ref{fig:nb}). These four fields span a range in redshift from $z_{\rm QSO}\approx 0.5$ to $z_{\rm QSO}\approx 1.1$, constituting the first $z\gtrsim 0.5$ sample with two-point characterisations of the CGM velocity field. While all fields host a bright QSO with a bolometric luminosity of $\sim 10^{47}$ erg s$^{-1}$, these QSOs span a range in radio luminosity and reside in diverse group environments with different numbers of neighboring galaxies found (see Table~\ref{tab:summary_table} for a summary of the QSO properties).  We have measured the second- and third-order VSFs over a range of scales, from $\lesssim 5$ kpc to $\approx 20$-50
kpc in these nebulae.  While no constraints on the energy injection and dissipation scales can be obtained from the current data,
we are able to determine a robust power-law slope after accounting for the effects of atmospheric seeing, spatial smoothing, and
large-scale bulk flows.  This work represents the first empirical study to resolve the turbulent velocity field in the CGM beyond the nearby Universe.

\begin{table}
\centering
\caption{Summery of the QSO properties.}
\label{tab:summary_table}
\begin{threeparttable}
\begin{tabular}{lcrccc} 
		\hline
		 &  &  & $\sigma_{v,{\rm group}}^{b}$& 
		 Radio \\ 
		Field name & Redshift & $N_{\rm group}^{a}$ & (km/s) & 
		mode \\ 
		\hline
		PKS0454$-$22$^{c}$ & 0.5335 & 23 & $\approx 320$& 
		Loud \\ 
		J0454$-$6116$^{d}$ & 0.7861 & 18 & $\approx 300$ & 
		Quiet \\ 
		J2135$-$5316 & 0.8115 & 2 & -- & 
		Quiet \\  
		TXS0206$-$048$^{e}$ & 1.1317 & 27 & $\approx 550$ & 
		Loud \\  
		\hline
	\end{tabular}
	\begin{tablenotes}
    \footnotesize
    \item \textbf{Notes.}
    \item[{\it a}] Number of spectroscopically-identified group member galaxies.
    \item[{\it b}] Velocity dispersion of the group.
    \item[{\it c}] QSO properties of PKS0454$-$22 are adopted from \cite{Helton2021}. While the authors   identified 23 galaxies with $|\Delta v|<1500$ km/s 
    and $d\lesssim 300$ kpc from the QSO location, the velocity distribution of these galaxies is clearly asymmetric with a tail extending to $\approx 1500$ km/s.  The velocity dispersion referenced here is calculated using 19 galaxies with $|\Delta\,v|<1000$ km/s.  
    \item[{\it d}] For both J0454$-$6116 and J2135$-$5316, group member galaxies are found with $|\Delta v|<1000$ km/s from the QSO redshift and $d\lesssim 250$ kpc from the QSO location 
    (J. Li, privare communication).
    \item[{\it e}] For TXS0206$-$048, group member galaxies are found with $|\Delta v|<1500$ km/s from the QSO redshift and $d\lesssim 500$ kpc from the QSO location 
    \citep[][]{Johnson2022}.
    \end{tablenotes}
    \end{threeparttable}
\end{table}

This paper is organized as follows.  First, we illustrate the basic formalism of VSFs in \S\,\ref{sec:VSF_bkg}, and discuss how the smoothing and projection effects in observational data can affect the VSF measurement. In \S\,\ref{sec:obs_and_measurement}, we present the IFS data used in this work, the subsequent emission line analyses, as well as the VSF measurements. The results are presented in \S\,\ref{sec:results}.  We discuss our results in \S\,\ref{sec:discussion}, and conclude in \S\,\ref{sec:conclusion}.  Throughout this paper, we adopt a Hubble constant of $H_0=70$ km/s/Mpc, $\Omega_\mathrm{M}=0.3$ and $\Omega_\Lambda = 0.7$ when deriving distances, masses and luminosities.  All distances quoted are in physical units. 

\section{Velocity structure functions as a tracer of turbulence}
\label{sec:VSF_bkg}
As defined in Equation~\ref{eq:vsf}, the VSF quantifies the kinetic energy fluctuations as a function of scale in a velocity field.  \cite{Kolmogorov1941} showed that for isotropic, homogeneous, and incompressible flows with sufficiently large Reynolds numbers, the VSF should follow a power-law scaling of $S_p(r)\propto r^{p/3}$.
In particular, the second-order VSF $S_2(r)\propto r^{2/3}$ is directly related to the auto-correlation function $\Gamma (r)$ and the kinetic energy power spectrum $E_k$ of an isotropic velocity field through
\begin{equation}
    S_2(r)=2[\Gamma(0) - \Gamma(r)] = 2\int (1-e^{ikr})E_kdk,
	\label{eq:S2}
\end{equation}
where 
\begin{equation}
    \Gamma(r)=\langle \vect{v}(\vect{x}) \vect{v}(\vect{x}+\vect{r}) \rangle
\end{equation} and $k=2\pi/r$.  The energy power spectrum then scales with $k$ following $E_k\propto k^{-5/3}$. Similarly, the third-order VSF $S_3(r)\propto r$ can be derived exactly to follow $S_3(r)=-(4/5)\,\epsilon\,r$, where $\epsilon$ represents the energy cascade rate (also see \S\,\ref{sec:k41} below for a discussion on $\epsilon$).

While these theoretical expectations of VSFs are established in three dimensional space, empirical data are 
limited to projected quantities. Specifically, the velocity differences are measured along the line of sight based on the observed Doppler shifts, 
and only projected separations $r_{\rm proj}$ along the plane of the sky are accessible instead of the true three-dimensional distances between two locations. Such limitations need to be accounted for explicitly when interpreting observational results. 

The effect of projections in the observed VSFs has been investigated extensively by previous authors.
When viewing a cloud with well-established three-dimensional Kolmogorov turbulence in projection, \cite{vonHoerner1951} demonstrated that the shape of the measured VSF depends on the depth, $L$, of the cloud along the line of sight.  At separations $r_{\rm proj}<L$, the VSF is expected to steepen, with a power-law slope of $5p/6$, but it recovers to the theoretical value of $p/3$ at larger separations 
$r_{\rm proj}>L$. There is a smooth transition between the two regimes that could be used as a probe of the cloud depth $L$ \citep{vonHoerner1951}. This effect, sometimes referred to as ``projection smoothing'', is also verified by several other studies both analytically \citep[e.g.][]{Odell1987, Xu2020} and in numerical simulations \citep[e.g.,][]{Mohapatra2022}. Meanwhile, a recent study by \cite{Zhang2022} suggests that if the emission source is more spatially-confined (e.g., H$\alpha$ filaments at the center of some galaxy clusters), the projection effect will flatten the VSF as opposed to making it steeper. 

In addition to line-of-sight projection effects, the spatial correlation due to atmospheric seeing in ground-based data 
will also alter the shape of the measured VSF.  Additional spatial smoothing often applied to enhance the signal-to-noise ratio (SNR) of noisy data would further increase the scale of the spatially-correlated signal. Fortunately, this effect can be analytically incorporated into the theoretical models of the second-order VSF $S_2$, allowing a more accurate comparison between data and model expectations. Based on Equation \ref{eq:S2}, the second-order VSF of a spatially-smoothed velocity field can be written as 
\begin{equation}
    S_2^\prime(r) = 2[\Gamma^\prime(0)-\Gamma^\prime(r)].
	\label{eq:S2_prime}
\end{equation}
$\Gamma^\prime(r)$ is the auto-correlation function of the smoothed velocity field and can be calculated by
\begin{equation}
    \Gamma^\prime(r) = \langle \vect{v}^\prime(\vect{x}) \vect{v}^\prime(\vect{x}+\vect{r}) \rangle=\vect{v}^\prime\,\otimes\,\vect{v}^\prime,
	\label{eq:Gamma_prime}
\end{equation}
where $\vect{v}^\prime$ is the smoothed velocity field.  If we designate $\vect{g}(\vect{x})$ as the spatial smoothing kernel, then the smoothed velocity field can be expressed as convolution of $\vect{v}$ with a Gaussian kernel representing the total point-spread-function (PSF), $\vect{v}^\prime(\vect{x})=\vect{g}(\vect{x})\ast \vect{v}(\vect{x})$. 
Equation \ref{eq:Gamma_prime} can now be rewritten as 
\begin{equation}
    \Gamma^\prime(r) = \vect{v}^\prime\,\otimes\,\vect{v}^\prime = (\vect{g}\ast\vect{v})\,\otimes\,(\vect{g}\ast\vect{v}) 
	\label{eq:Gamma_prime2}
\end{equation}
Equation \ref{eq:Gamma_prime2} can be rearranged to a simple analytic form of
\begin{equation}
    \Gamma^\prime(r) = (\vect{g}\,\otimes\,\vect{g}) \ast (\vect{v}\,\otimes\,\vect{v})=\Gamma_g(r)\ast \Gamma(r).
	\label{eq:Gamma_prime_2}
\end{equation}
Equation \ref{eq:Gamma_prime_2} shows that the auto-correlation function of a smoothed velocity field can be calculated through a convolution of two functions: the auto-correlation function of the smoothing kernel, and the auto-correlation function of the intrinsic, unsmoothed velocity field.  When both $\Gamma_g(r)$ and $\Gamma(r)$ have analytical expressions, such as the case for a Gaussian smoothing kernel and a power-law auto-correlation function, $\Gamma^\prime(r)$ can be calculated explicitly and an exact expression for $S_2^\prime$ can be obtained through Equation \ref{eq:S2_prime}. 

\begin{figure}
	\includegraphics[width=\linewidth]{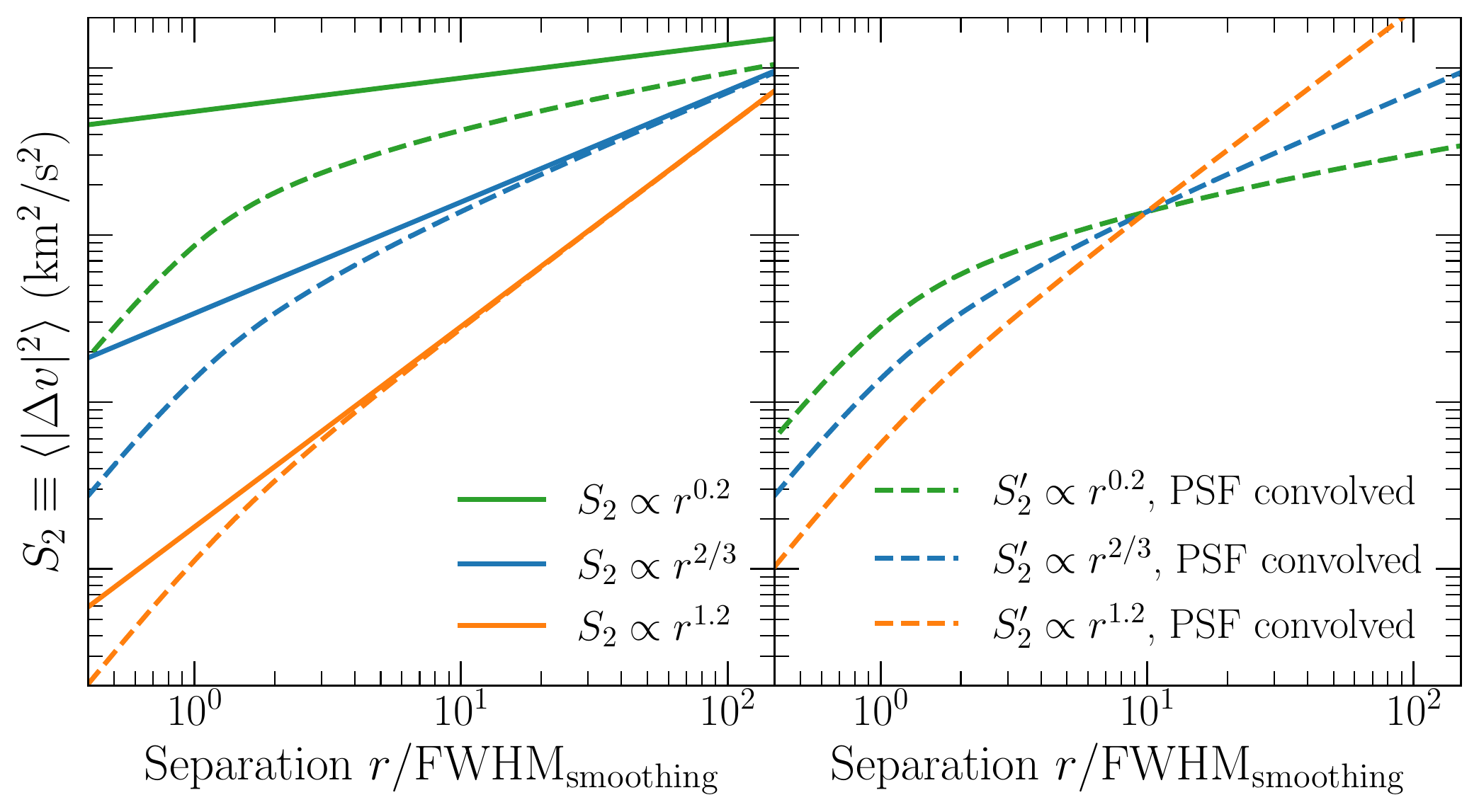}
    \caption{{\it Left}: Illustration of the spatial smoothing effect on the shape of the second-order VSF.  Green, blue and orange solid lines show power-law $S_2(r)$ with an intrinsic slope of $\gamma_2=0.2$, 2/3 (i.e., Kolmogorov slope) and 1.2, respectively. The corresponding dashed curves show the shapes of $S^{\prime}_2(r)$ after convolving with a Gaussian smoothing kernel, calculated with Equations \ref{eq:S2_prime}--\ref{eq:Gamma_prime_2}.  It can be seen that spatial smoothing significantly steepens the VSF at $r\lesssim 2\times {\rm FWHM_{smoothing}}$, and the discrepancy is stronger for a flatter intrinsic VSF, as discussed in the text. {\it Right}: Smoothed $S^{\prime}_2(r)$ curves, same as shown in the left column, re-normalised to the same value at $r=10\times {\rm FWHM_{smoothing}}$.  This panel shows that with an accurate estimate of the smoothing kernel size, the intrinsic VSF slope can be obtained with high-SNR measurements even if the probed spatial scale does not cover a large dynamic range. } 
    \label{fig:model_w_3slopes}
\end{figure}

To visualize this spatial smoothing effect, we perform a series of calculations, adopting three different intrinsic power-law slopes for $S_2$, corresponding to a relatively flat VSF with a slope of $\gamma_2=0.2$, a Kolmogorov VSF of $\gamma_2=2/3$, and a steeper VSF of $\gamma_2=1.2$.  Assuming a Gaussian kernel for spatial smoothing, the comparisons of the intrinsic $S_2$ and the smoothed $S_2^\prime$ are shown in Figure \ref{fig:model_w_3slopes}.  It is clear that the smoothing effect is more significant for a flatter intrinsic VSF.  This can be understood intuitively by noting that a flatter VSF carries significantly more relative power on small scales, corresponding to large $k$ modes.
As a result, spatial smoothing, which by design removes the power from large $k$ modes, will have a more significant impact in systems with a flatter energy power spectrum.  Taking the Kolmogorov VSF for reference, Figure \ref{fig:model_w_3slopes} shows that the measured VSF begins to recover the intrinsic, unsmoothed VSF at separations $\gtrsim 4$ times the full-width-at-half-maximum (FWHM) of the smoothing kernel.  We, therefore, emphasize the importance of explicitly taking into account this smoothing effect in VSF measurements, especially when working with data where the seeing size is relatively large compared with the scales probed. 

For the QSO nebulae included in the current study, the spatial scales probed are restricted to $\lesssim 10$ times the FWHM of the PSF (see \S\,\ref{sec:obs_and_measurement} below). Fortunately, as we show in the right-hand panel of Figure \ref{fig:model_w_3slopes}, with sufficient SNR in the VSF measurements and an accurate estimate of the PSF size, the intrinsic VSF slope can still be recovered even when working with a limited dynamic range. Similarly, for the VSF measurements of H$\alpha$ filaments in cluster cores \citep{Li2020}, the steeper slopes may be partially attributed to the spatial smoothing effect due to the limited dynamic range compared to the size of the seeing disk in the data. 


We have demonstrated that it is straightforward to incorporate any spatial smoothing present in the data to the second-order VSF measurements, thanks to the convenient relation between $S_2$ and the auto-correlation function $\Gamma(r)$.  It is less straightforward to do so for the third-order VSFs, from which we expect to infer the energy cascade rate based on the exact relation of $S_3(r)\propto \epsilon\,r$ (see the discussion in \S\,\ref{sec:k41} below).  \cite{Benzi1993} reported the existence of an extended self-similarity (ESS), where VSFs of different orders are tightly correlated with each other and roughly following a simple power-law function.
The ESS is useful because it applies to cases with both high and low Reynolds numbers.  For example, for cases with low Reynolds numbers,  
the second-order VSFs 
may not follow the expected power-law scaling relation due to a lack of a well-established inertial range.  However, with ESS, the third-order VSFs can still be inferred to constrain the energy cascade rate.  In addition, with a simulated velocity field generated using Fourier series \citep[see e.g.][]{Saad2017}, we have tested that the spatial smoothing effect does not alter the power-law scaling relation for ESS. In other words, if a velocity field exhibits an ESS relation of $S_{p}=\alpha S_{3}^{\gamma_p/\gamma_3}$, then this scaling relation is preserved as $S^{\prime}_{p}=\alpha S^{\prime\,\gamma_p/\gamma_3}_{3}$ after the velocity field is smoothed. 
In \S\,\ref{sec:VSF_of_txs0206} \& \S\,\ref{sec:ESS_discussion} below, we show that the ESS is observed in all systems and discuss the caveats associated with this observation.  

\section{Observations and measurements}
\label{sec:obs_and_measurement}

To measure the VSFs in extended nebulae, spatially-resolved velocity maps are necessary.  In this section, we described the wide-field IFS observations available for detecting extended nebulae around four QSO hosts and the constructions of velocity maps based on line profile analyses of \oii and  \oiii emission lines.  

\subsection{IFS Observations}
\label{sec:data} 

\begin{table}
	\centering
	\caption{Journal of MUSE observations.}
	\label{tab:QSO_obs_info}
	\begin{tabular}{lccrc} 
		\hline
		 & & & $t_{\rm exp}$ & seeing$^a$ \\
		Field name & RA(J2000) & Dec.(J2000) & (s) & ($''$)  \\
		\hline
		PKS0454$-$22 & 04:56:08.90 & $-$21:59.09.1 & 2700 & 0\farcs6 \\
		J0454$-$6116 & 04:54:15.95 & $-$61:16:26.6 & 5100 & 0\farcs7 \\
		J2135$-$5316 & 21:35:53.20 & $-$53:16:55.8 & 6840 & 0\farcs6 \\
		TXS0206$-$048 & 02:09:30.74 & $-$04:38:26.5 & 28800 & 0\farcs7\\
		\hline
	\end{tabular}
	\begin{tablenotes}
    \footnotesize
    \item \textbf{Notes.}
    \item $^{\it a}$ Atmospheric seeing FWHM measured using the QSO at 7000\ang.  To improve the quality of line fitting, each combined data cube was convolved with a Gaussian kernel of FWHM$=0\farcs7$. This yielded a total PSF FWHM of $\approx 0\farcs9$-$1\farcs0$ (see \S\,\ref{sec:nb_3dmask}), corresponding to a projected separation of 6-8 kpc at the redshifts of these QSOs.
    \end{tablenotes}
\end{table}

Wide-field IFS data of the QSO fields 
were obtained using the Multi-Unit Spectroscopic Explorer (MUSE; \citealp{Bacon2010}) on the VLT UT4. All four fields were observed under the Wide-Field-Mode (WFM), which provides a contiguous field-of-view (FOV) of $1\arcmin\times1\arcmin$ in a single pointing, with $0\farcs2$ per pixel spatial sampling. MUSE covers a wavelength range of 4750--9350  $\ang$ with a resolving power of $R\approx 2000$--4000 (higher at the longer wavelength end). 

\begin{table*}
	\centering
	\caption{Summery of emission properties in spatially-extended QSO nebulae$^a$.}
	\label{tab:emission_line_properties}
    \begin{threeparttable}
	\begin{tabular}{lcccccccc} 
		\hline
		 & \multicolumn{2}{c}{Surface Brightness Limit$^{b}$} & & \multicolumn{2}{c}{Luminosity (erg s$^{-1}$)} & & \multicolumn{2}{c}{Nebula area (kpc$^2$)} \\
		 \cline{2-3}\cline{5-6}\cline{8-9} 
		Field name &[\ion{O}{II}] &[\ion{O}{III}] & & [\ion{O}{II}] &[\ion{O}{III}] & & [\ion{O}{II}]& [\ion{O}{III}] \\
		\hline
		PKS0454$-$22 & $2.3\times 10^{-19}$ & $1.7\times 10^{-19}$ & & $1.9\times 10^{42}$ & $2.2\times 10^{43}$ & &  1552 & 2202 \\
		J0454$-$6116 & $1.2\times 10^{-19}$ & $2.4\times 10^{-19}$ & & $3.5\times 10^{42}$ & $5.3\times 10^{42}$ & & 3821 & 2128 \\
		J2135$-$5316 & $1.4\times 10^{-19}$ & $2.6\times 10^{-19}$ & & $2.5\times 10^{42}$ & $9.2\times 10^{42}$ & & 1614 & 2190 \\
		TXS0206$-$048 & $6.3\times 10^{-20}$ & -- & & $2.0\times 10^{43}$ & -- & & 6239 & -- \\
		\hline
	\end{tabular}
	\begin{tablenotes}
	\footnotesize
    \item \textbf{Notes.}
    \item[{\it a}]  Luminosities and nebula sizes are summed over the areas used for the subsequent VSF analysis, which are smaller than the areas shown in Figure \ref{fig:nb}. See velocity maps (e.g. Figure \ref{fig:TXS0206_w_grad}) for the areas included in the VSF calculation. Note that for the nebula in TXS0206$-$048, the $r<1\arcsec$ region centered on the QSO contributes to $\approx 50\%$ of the total luminosity.  Excluding this central region results in a luminosity estimate consistent with the reported value in \cite{Johnson2022}. 
    \item[{\it b}] 1-$\sigma$ limit in units of $\sbunit$ over a single wavelength slice (i.e., 1.25$\ang$) at the observed wavelength of the corresponding emission line.
    \end{tablenotes}
    \end{threeparttable}
\end{table*}

Out of the four fields, J0454$-$6116 and J2135$-$5316 were obtained as part of the Cosmic Ultraviolet Baryon Survey (CUBS) using adaptive optics assisted WFM under program ID, 0104.A-0147 \citep[PI: H.-W. Chen;][]{CUBS1}. The total exposure time was 5100s for J0454$-$6116 and 6840s for J2135$-$5316. PKS0454$-$22 was observed under program ID 0100.A-0753 \citep[PI: C. P\'eroux;][]{Peroux2019}, with a total exposure time of 2700s.  TXS0206$-$048 was part of the MUSE Quasar-field Blind Emitters Survey (MUSEQuBES) under program IDs 097.A-0089(A) and 094.A-0131(B) \citep[PI: J. Schaye;][]{Muzahid2020} with a total exposure time of 28,800s.  All observations were carried out under good seeing conditions, with the mean seeing FWHM measured to be  $\approx0\farcs6$--$0\farcs7$ at the location of the QSOs at 7000$\ang$.  A summary of the MUSE observations, including the mean seeing in the final combined data cube, is listed in Table~\ref{tab:QSO_obs_info}.

Raw science exposures and the associated raw calibration files were retrieved from the ESO science archive. We reduced the data of all four fields using the standard ESO MUSE pipeline \citep[v.2.8.4;][]{Weilbacher2020}, and applied an additional sky subtraction in the final combined cubes using the median sky spectrum obtained from object-free regions in each field. 

The pipeline-generated variance cube has been known to underestimate the data uncertainties \citep[e.g.][]{Bacon2017}.  Using the wavelength range of 6000--7000$\ang$, we obtained an empirical estimate of the uncertainties in each field and found that on average this empirical noise level is $\approx 1.6$ times higher than the noise level inferred from the pipeline generated variance cube. We, therefore, scaled up the pipeline-produced variance cube by a factor of 1.6$^2$. This correction factor is similar to what has been adopted in previous studies \citep[e.g.][]{Borisova2016,Sanderson2021}.

\subsection{QSO light subtraction}
\label{sec:QSO_subtraction} 
To better reveal the emission from the extended nebulae, we removed the QSO light following a method similar to the high-resolution spectral differential imaging technique discussed in \cite{Haffert2019} and \cite{Xie2020}. Below we briefly describe the main steps.

We first constructed a QSO template spectrum using the mean spectrum from the central 5 spaxels (i.e., within a radius of $0\farcs2$) around the QSO. Next, for each spaxel contaminated by the QSO light, we divided the data in this spaxel by the QSO template spectrum to obtain a ratio spectrum.  We then smooth this ratio spectrum with a median rolling filter with a window width of $\sim 100$ spectral pixels (i.e., $\sim125 \ang$). The exact window size is decided through trial and error and is slightly different for different fields. This smoothing step will maintain the low-order variation in the ratio spectrum while removing high-order features, such as strong emission lines and noise. Finally, we scale the QSO template spectrum by the smoothed ratio spectrum and subtract it from the spaxel to remove the QSO contamination. These steps were repeated for every spaxel within a radius of 30 pixels (i.e., 6\arcsec) from the QSO center in each field. 

Comparing with other commonly used QSO light subtraction methods, such as principle component analysis (and similarly, non-negative matrix factorization) \citep[e.g.][]{Johnson2018, Helton2021} and a joint analysis of the QSO spectrum and the host galaxy spectrum incorporating stellar population synthesis models \citep[e.g.][]{Rupke2017}, the method described above has the advantage of being relatively simple while delivering very clean residual spectra. However, a couple of caveats should also be noted. By using a QSO template spectrum that is scaled according to the smoothed ratio spectrum, this method removes all low-order features, including continuum and broad emission lines, indiscriminate to the origin of such features. As a result, it removes the low-order signal from the QSO host galaxy as well as other possible continuum sources located underneath the QSO PSF. Hence this method works well for revealing spectral features narrower than typical QSO broad lines, such as the extended nebula emission studied here, but it is not suitable for studies of host galaxies and continuum sources. Meanwhile, in the QSO template spectrum constructed around the core region of the QSO PSF, there are possible contributions from the template to the targeted narrow emission line, and therefore the line flux in the nebula after QSO light subtraction could be underestimated. We, therefore, take extra caution when forming the QSO template and exclude spaxels with relatively strong narrow signals at the wavelength of the lines of interest. 

\subsection{Narrow-band images}
\label{sec:nb_3dmask} 
Additional continuum subtraction was applied to the full data cube across the FOV to further remove background continuum flux in spaxels not included in the QSO light subtraction step. In general, we used a continuum spectrum determined through linear interpolation using the median value in a blue (red) window that was approximately [$-$3000, $-$1500] ([$+$1500, $+$3000]) km/s away from the expected line center. In practice, based on the observed wavelength of the line of interest in each field, the spectral windows were adjusted to avoid noisy regions due to strong skylines and other artifacts in the data cube. 

To enhance the SNR of the extended faint emission in the outskirts of each QSO nebula, we smoothed the data in the spatial dimension with a Gaussian kernel. The FWHM of the Gaussian kernel is chosen to be $3.5$ pixels (i.e., $0\farcs7$) for all four fields.  In Table~\ref{tab:QSO_obs_info}, we list the atmospheric seeing size for each field measured at the position of the QSO at 7000$\ang$ before applying the additional spatial smoothing.  The total PSF FWHM after smoothing was $\approx 50$\% larger than the seeing disk. No additional smoothing was applied along the spectral dimension.  The 1-$\sigma$ surface brightness limit in a single wavelength slice (i.e., width of 1.25$\ang$) at the observed wavelengths of the \oii and \oiii lines for each field ranges from approximately $6\times 10^{-20}\sbunit$ to $3\times 10^{-19}\sbunit$, as listed in Table~\ref{tab:emission_line_properties}.  TXS0206$-$048 has the lowest noise level at the observed \oii wavelength due to a significantly longer total integration time.

With the smoothed, continuum- and QSO light-subtracted data cube, optimally-extracted narrow-band images were constructed for both \oii and \oiii lines for the three lower redshift fields.  For TXS0206$-$048, the \oiii line is redshifted out of the MUSE spectral coverage and therefore no results based on \oiii are available.  Optimal extraction using a 3D mask that takes advantage of the information along both the spatial and the spectral dimension was carried out to maximize the SNR in the resulting narrow-band images.  Detailed descriptions of the procedure can be found in, e.g.,  \cite{Borisova2016} and \cite{Sanderson2021}. In short, the 3D mask was created based on an SNR threshold chosen for each voxel (i.e., volume pixel) of the data cube, and the minimum number of consecutive spectral pixels in a given spaxel to be included in the mask.  For this study, we chose a voxel SNR threshold of 1 and a minimum number of consecutive spectral pixels of 3. The narrow-band images constructed using such 3D masks are shown in Figure \ref{fig:nb} for all four fields. 

\subsection{Emission line analysis and velocity measurements}
\label{sec:line_fitting} 
To determine the line-of-sight velocities at different locations across the nebulae, we performed a line profile analysis by adopting 
a Gaussian profile convolved with an appropriate instrumental line spread function. For fields with both \oii and \oiii detections, we carried out the analysis of these two lines separately. The \oii doublet is mostly unresolved for all fields at the MUSE resolution. Furthermore, both the \oii and \oiii emission lines exhibit complex line profiles showing evidence for multiple velocity components, particularly in the inner regions closer to the QSOs. For some individual velocity components, spatial variation is observed in the [\ion{O}{III}]/[\ion{O}{II}] line ratio, leading to different flux-weighted mean velocities at the same locations for these two lines.  We, therefore, decided to take a simpler approach and fit these two lines separately. 

We adopted an MCMC approach to search for the best-fit parameters of individual Gaussian components, which was implemented with the Python module \texttt{emcee} \citep[][]{emcee}. Compared with least-square-based fitting methods, an MCMC approach provides a more robust posterior probability density distribution for the model parameters, naturally accounting for non-Gaussian posteriors as well as upper/lower limits. For the \oii blended doublet, we only included one Gaussian component in the model profile for all spaxels, as the current MUSE data do not provide sufficient spectral resolution to break the degeneracy between the centroids of multiple velocity components and the doublet line ratios. For the \oiii line, we conducted the fitting with up to four independent Gaussian components and determined the number of components in each spaxel based on the Bayesian information criterion (BIC) \citep[see e.g.][]{mcmc_review_sharma2017}.  We required that a complex model with more Gaussian components can be accepted only when its BIC value was smaller than the BIC value of a simpler model by at least 30. We chose this more stringent threshold than the commonly adopted value of $\Delta {\rm BIC}>10$ because the spectra from the data cubes often displayed complicated noise spikes that were not fully accounted for in the error arrays, and a more conservative approach was required to avoid over-fitting when using multiple components. 

However, as we will discuss in \S\,\ref{sec:VSF_other_3_fields} and show in Figure \ref{fig:o3_multicomp_compare} below, for spaxels with multiple Gaussian components to model the observed \oiii line, adopting a flux-weighted mean velocity leads to similar VSF measurements as adopting a one-component model (i.e., ignoring the multi-component nature of the line).  For simplicity, we, therefore, opted to focus on the VSF measurements based on the one-component model even for spaxels with complex line profiles. We will present and discuss results from the multi-component fitting process of the \oiii line in a subsequent paper.


\subsection{VSF measurements}
\label{sec:VSF_measurement} 
Before carrying out the VSF measurements, we performed a series of checks to ensure that the results are robust.
First, we examined possible contamination resulting from overlapping continuum sources due to projection effects.  
In particular, a large velocity contrast would suggest that such continuum sources might not belong to the same dynamic system as the rest of the line-emitting gas, and therefore should be excluded from the VSF measurements. For the PKS0454$-$22 and TXS0206$-$048 fields, we used the archival broadband {\it HST} data to identify continuum sources (\citealp{Helton2021}; Johnson et al.\ in prep).  For the J0454$-$6116 and J2135$-$5316 fields, due to a lack of higher spatial resolution imaging data, continuum sources were identified using a MUSE white-light image.  We flipped the white-light image of each field along the x-axis (using the QSO centroid as the center), and subtracted the flipped image from the original image. 
Strong continuum sources underneath the QSO PSF will lead to a pattern of significant residual flux at the original locations of such sources paired with significant over-subtraction at their flipped locations. This method helps to identify sources that might be easy to miss due to the QSO PSF.   Flipping along the y-axis of the image would have achieved the same effect. Out of the four fields, we only identified two strong continuum sources in the J0454$-$6116 field that stood out in the velocity map and excluded the spaxels inside a circular aperture centered on each of these two continuum sources.  The size of the aperture was chosen to enclose most of the continuum flux. For the remaining three fields, the continuum sources overlapping with the nebulae showed consistent velocities with the rest of the nebulae, and no spaxels were excluded from the VSF measurements. 

In addition, we masked spaxels with highly uncertain velocity measurements.  Because we adopted a generous voxel SNR threshold when forming the 3D masks (see \S\,\ref{sec:nb_3dmask}), some spaxels included in the line fitting step had relatively faint signals and large measurement uncertainties that would significantly impact the VSF measurement uncertainties. 
We excluded spaxels with a velocity uncertainty larger than 45 km/s.  This threshold was approximately two to three times the median uncertainty of the fitting results based on the \oiii line, and was about the median uncertainty for measurements based on the \oii line among all fields. We verified that changing this threshold by a small amount (i.e., $\pm 15$ km/s) did not lead to significant differences in the subsequent analyses. Finally, we examined the probability density distribution of the observed velocities among the rest of the spaxels in each field, and filtered out spaxels that are outliers (i.e., either too blue or too red in velocity, defined to be the $\approx 2$\% tail on both ends). We also excluded the central $r\le3$ pixels region for both J0454$-$6116 and J2135$-$5316 fields due to noisy residuals from the QSO light removal, which was not necessary for PKS0454$-$22 and TXS0206$-$048. 

\begin{figure*}
	\includegraphics[width=\textwidth]{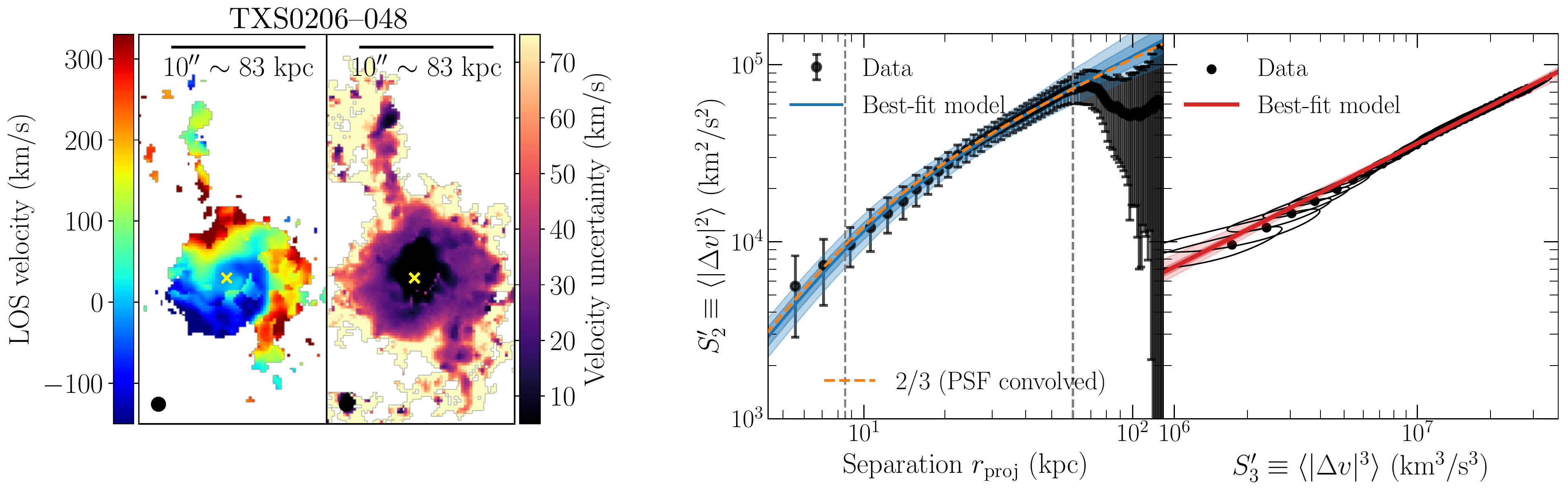}
    \caption{{\it Left panels:} The observed velocity map and the associated measurement uncertainties of the extended nebula around TXS0206$-$048 using the \oii emission lines. The yellow cross in both panels marks the QSO position. Only pixels included in the VSF calculation are shown in the velocity map (see \S\,\ref{sec:VSF_measurement}), while the velocity uncertainty panel contains all pixels from the 3D mask (see \S\,\ref{sec:nb_3dmask}). The black circle at the bottom left represents the total PSF of this field, after convolving the atmospheric seeing with the smoothing kernel applied to improve the SNR. {\it Right panels:} The second-order VSF $S^{\prime}_2(r)$ constructed using the velocity map displayed in the left panel, along with the $S^{\prime}_2$ vs.\ $S^{\prime}_3$ correlation. Vertical dashed lines mark the fitting boundaries in $r_{\rm proj}$, with the left line indicating the FWHM$_{\rm total}$ of the field and the right line indicating the maximum $r_{\rm proj}$ beyond which a single power-law model does not provide a good fit anymore. The best-fit model of $S^{\prime}_2$, after being convolved with the total PSF, is shown as the blue solid curve. The dark (light) blue shaded region represents the 16$^{\rm th}$--84$^{\rm th}$ (2$^{\rm nd}$--98$^{\rm th}$) quantile range for the model. The orange dashed curve shows the PSF-convolved Kolmogorov mode for $S^{\prime}_2$ with a theoretical slope of $\gamma_2=2/3$.  In the right panel, the best-fit power-law model for the $S^{\prime}_2$ vs.\ $S^{\prime}_3$ relation is shown as the red solid line with the model uncertainty represented by the red shaded region. Only the data points within the same distance separation range for the fitting of $S^{\prime}_2$ are shown in the $S^{\prime}_2$-$S^{\prime}_3$ panel, with the ellipses showing the correlated 1-$\sigma$ error area determined by the eigen vectors and eigen values of the covariance matrix within each distance bin.  We measure an intrinsic power-law slope of $\gamma_2={0.72^{+0.12}_{-0.11}}$ and $\gamma_3={1.03^{+0.18}_{-0.16}}$ for $S_2$ and $S_3$, respectively (see Table~\ref{tab:slopes}).}
    \label{fig:TXS0206_w_grad}
\end{figure*}

All spaxels left after the above filtering steps were included in subsequent VSF measurements.  Summing over all these spaxels, we report the total luminosity in \oii and \oiii line emission as well as the total area (in kpc$^2$) for each field in Table \ref{tab:emission_line_properties}. Out of the four fields, TXS0206$-$048 has the largest area.  In fact, in terms of the area and the total line emission luminosity of the [\ion{O}{II}] nebula, TXS0206$-$048 exceeds 
the ``Makani" nebula at $z=0.459$, the largest [\ion{O}{II}] nebula detected hitherto \citep[][]{Rupke2019}\footnote{Note that the size and [\ion{O}{II}] line luminosity of TXS0206$-$048 reported in Table \ref{tab:emission_line_properties} were obtained at a surface brightness level of $\sim 10^{-19}\sbunit$, significantly lower than the surface brightness threshold of $5\times 10^{-18}\sbunit$ used in \cite{Rupke2019}.  At the level of $5\times 10^{-18}\sbunit$, the TXS0206$-$048 [\ion{O}{II}] nebula has an area $\approx 4100$ kpc$^2$, slightly smaller than 4900 kpc$^2$ covered by ``Makani".}.  The filtered [\ion{O}{II}] velocity map of the TXS0206$-$048 field is shown in the left panel of Figure \ref{fig:TXS0206_w_grad}, together with its corresponding velocity uncertainty map on the whole 3D mask footprint for comparison. The filtered velocity maps of the other three fields, for both [\ion{O}{II}] and [\ion{O}{III}] lines, are shown in Figures \ref{fig:PKS0454_o2}--\ref{fig:J2135_o3} in the Appendix. 

Because of the spatial correlation between adjacent spaxels, when measuring the VSF, individual velocity pairs within a distance separation bin are not independent of each other.  We therefore cannot directly propagate the measurement uncertainties of the  velocity centroids in each spaxel to estimate the uncertainties of the VSFs. To robustly estimate the uncertainty of the VSF, we proceeded with the following steps.  First, we divided the whole nebula in each field into smaller sub-regions.  The size of these sub-regions was roughly the FWHM of the total PSF in each field (see Table \ref{tab:summary_table}). Most of these sub-regions were squares while some sub-regions located near the edge of the nebula had irregular shapes. Next, we randomly selected one spaxel per sub-region and constructed a VSF based only on the selected spaxels.  We then repeated the step of randomly selecting one spaxel per sub-region 1000 times, and each time obtained a VSF measurement.  In addition, for each iteration, we perturbed the velocity map to within the measurement uncertainties by randomly assigning a new velocity value drawn from the MCMC chain to each spaxel. 
By restricting the pair formation to one spaxel per sub-region defined by the PSF, we were able to minimize correlated noise between adjacent bins in the VSF and recover small-scale power lost due to smoothing.
We refer to this procedure as a modified bootstrap method.  We obtained a mean and standard deviation of the 1000 VSFs as the measurement and associated uncertainty of the final VSF.  Note that while all VSFs were measured using a distance bin size of one spaxel, only measurements separated by scales larger than the size of the total PSF were included when quantifying the slope of the VSFs (see \S\,\ref{sec:VSF_of_txs0206} below for details of constraining the VSF slopes). 

\section{Results}
\label{sec:results}
Of the four QSO nebulae studied here, TXS0206$-$048 has the most constraining IFS data and the largest spatial extent (see Figure \ref{fig:nb} and Table \ref{tab:emission_line_properties}).  Together, these characteristics ensure the best-determined velocity map and well-constrained VSFs.  In this section, we present the VSFs measured for extended QSO nebulae at $z_{\rm QSO}\approx0.5$--1.1 with a focus on the line-emitting gas detected around TXS0206$-$048 at $z_{\rm QSO}\approx1.1$.  In addition, we investigate the impact on the observed VSFs due to possible underlying coherent bulk flows in these nebulae.  We consider the presence of unidirectional velocity gradient, radial, and tangential motions in the observed velocity field of each nebula, and compare the measured VSFs before and after removing these smooth velocity components.  

\subsection{The observed VSFs of TXS0206$-$048}
\label{sec:VSF_of_txs0206}

The velocity and velocity uncertainty maps of the \oii nebula around TXS0206$-$048 displayed in Figure \ref{fig:TXS0206_w_grad} show that the line-emitting gas is highly disturbed with well-determined line-of-sight velocities spanning a wide range from $\approx\,-150$ km/s to $\gtrsim\,300$ km/s across the full extent of nearly 200 kpc defined by the narrow stream-like feature toward the northeast and southwest \citep[][]{Johnson2022}.  However, most of the statistical power in the VSF measurements lies in the main, more spherically distributed nebula of $\approx\,90$ kpc in diameter centered on the QSO.  The observed second-order VSF, $S^{\prime}_2$, is well characterized by a single power-law scaling up to $r_{\rm proj}\approx\,60$ kpc over the projected distance range from $r_{\rm proj}<6$ kpc to $r_{\rm proj}\approx\,60$ kpc (Figure \ref{fig:TXS0206_w_grad}).

To quantify the second-order VSF slope, we apply a power-law model convolved with the total PSF to characterize the reconstructed $S^{\prime}_2$ from each of the 1000 realizations obtained through the modified bootstrap method described above.  We adopt a Gaussian function with an FWHM of 8.3 kpc for the PSF in TXS0206$-$048 (see Table \ref{tab:summary_table}), and we follow the steps discussed in \S\,\ref{sec:VSF_bkg} to calculate the shape of the power-law model after the PSF convolution. Note that we only consider non-negative power-law slopes, as negative slopes are not motivated by the data here and would lead to divergence at $r=0$ for a simple power-law parameterization.  The model fitting is done over the distance range of 8.3 kpc$<r_{\rm proj} <$60 kpc, using the \texttt{Scipy} curve\_fit routine. 
The small-scale cutoff at 8.3 kpc is to minimize systematic uncertainties due to spatial smoothing, while the large-scale threshold at 60 kpc is determined based on a series of trials and errors to optimize the fitting precision and accuracy. 
For $S_2$ of TXS0206$-$048 obtained using the \oii emission line, we measure a slope of $\gamma_2=0.72^{+0.12}_{-0.11}$. The best-fit value corresponds to the median value among the 1000 fitting results, and the 16$^{\rm th}$ and 84$^{\rm th}$ quantiles represent the lower and upper limit, respectively.  

In Figure \ref{fig:TXS0206_w_grad}, the best-fit model is shown in the blue solid curve, with the dark (light) blue shaded region representing the 16$^{\rm th}$--84$^{\rm th}$ (2$^{\rm nd}$--98$^{\rm th}$) quantile range for the models. This measurement is consistent with the Kolmogorov slope of $2/3$ (orange dashed curve in Figure \ref{fig:TXS0206_w_grad}) for isotropic, homogeneous, and incompressible turbulence. We have experimented with removing the stream-like features both north- and south-ward of the main nebula, and we obtained consistent VSF measurements.   

In addition to $S^{\prime}_2$, for each of the 1000 modified bootstrap samples described above, we also calculate the VSF $S_p^{\prime}$ for other orders up to $p=6$, and examined if the ESS discussed in \S\,\ref{sec:VSF_bkg} applies to this data set. Limited by the data quality, VSFs for $p>6$ become too noisy to result in meaningful constraints. 
In the right-most panel of Figure \ref{fig:TXS0206_w_grad}, we show the measurement of $S_2^{\prime}$ as a function of $S_3^{\prime}$ for TXS0206$-$048. Note that the $S^{\prime}_2$ and $S^{\prime}_3$ measurements are highly correlated.  Therefore, we use ellipses to show the 1-$\sigma$ confidence intervals with the 
elongations and sizes determined by the eigen vectors and eigen values of the data covariance matrix in each distance bin.

Similar to the ESS presented in \cite{Benzi1993} (see their Figure 3), we observe a well-defined power-law relation of $S^{\prime}_2\propto S^{\prime\,0.70\pm0.03}_3$. The measurement of this power-law slope is obtained using the 1000 realizations of the velocity map, and only the data points within the same distance range of 8--60 kpc are included in the model fitting. Due to the tight correlation between $S^{\prime}_2$ and $S^{\prime}_3$, the ESS scaling slope is much better constrained than the individual slopes $\gamma_2$ and $\gamma_3$. Because we can analytically incorporate the effect of PSF smoothing into a power-law $S_2$ but not for $S_3$, the presence of the ESS in this data set conveniently allows us to measure a slope and amplitude of $S_3$ accurately.  In addition, as discussed in \S\,\ref{sec:VSF_bkg}, the smoothing effect does not change the power-law scaling relation for ESS. 
Combining the measured $S_2$ slope of $\gamma_2=0.72^{+0.12}_{-0.11}$ and the $S_2$-$S_3$ power-law scaling of $\gamma _2/\gamma _3=0.70\pm0.03$, we obtain a slope of $\gamma_3=1.03^{+0.18}_{-0.16}$ for $S_3$ in TXS0206$-$048.  Consistent with the result for $S_2$, the $S_3$ slope is in excellent agreement with the expectation of $\gamma_3=1$ for Kolmogorov turbulence.  Discussions on the slopes of higher-order VSFs are presented in \S\,\ref{sec:ESS_discussion}. 

\subsection{Effect of large-scale velocity gradients} 
\label{sec:1Dgrad}

While the measured $S_2$ and $S_3$ are both consistent with Kolmogorov turbulence for the nebula surrounding TXS0206$-$048, a caveat remains regarding the presence of large-scale coherent flows which could contribute to the observed power in the velocity structure functions \citep[e.g.][for a discussion]{Zhang2022}.  In this section, we address this issue by considering a unidirectional flow model for removing the bulk flow in the observed velocity map.  

\begin{figure*}
	\includegraphics[width=\textwidth]{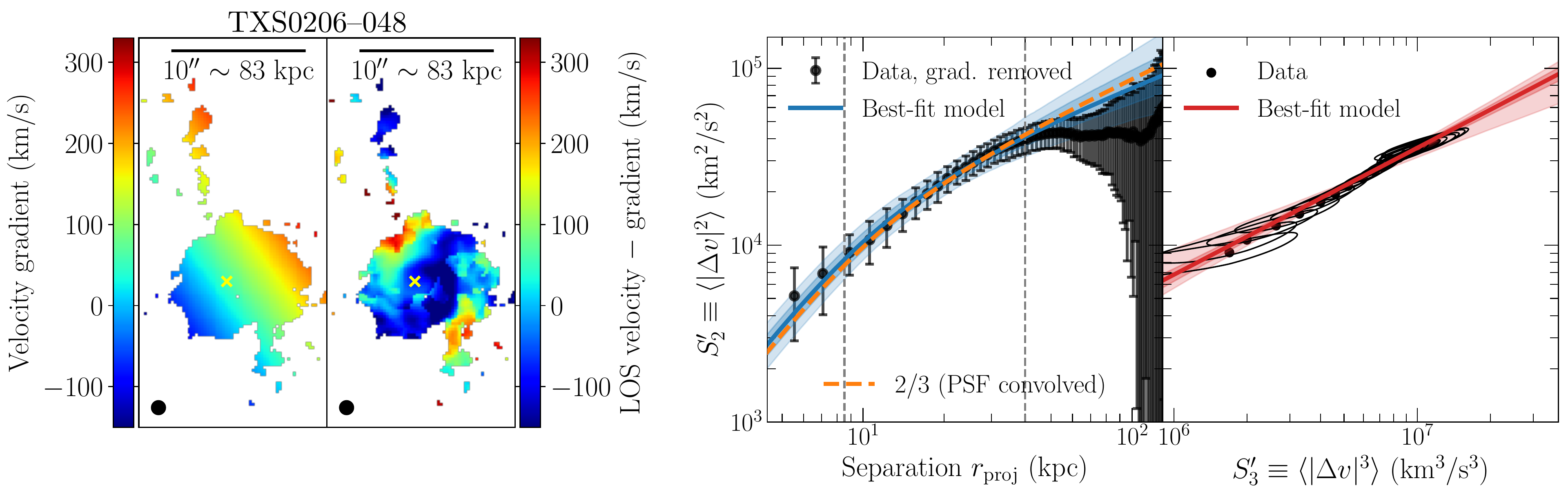}
    \caption{{\it Left panels:} The best-fit 2D velocity gradient model and the residual velocity map after subtracting the velocity gradient from the observed velocity map for TXS0206$-$048. The yellow cross in both panels marks the QSO position. Similarly to the left-most panel of Figure \ref{fig:TXS0206_w_grad}, only pixels included in the VSF calculation are shown. {\it Right panels:} $S^{\prime}_2(r)$ and the $S^{\prime}_2$-$S^{\prime}_3$ ESS relation constructed using the gradient-subtracted velocity map displayed on the left. Similar to the right panels of Figure \ref{fig:TXS0206_w_grad} but calculated with the gradient removed velocity map. We measure an intrinsic power-law slope of $\gamma_2={0.56^{+0.16}_{-0.17}}$ and $\gamma_3={0.78^{+0.28}_{-0.25}}$ for $S_2$ and $S_3$, respectively (see also Table ~\ref{tab:slopes}).}
    \label{fig:TXS0206_wo_grad}
\end{figure*}

We first adopt a simple model velocity map parameterized as $v(x,y)=ax+by+c$, where $x$ and $y$ are the coordinates of individual spaxels within the nebula, and $a$, $b$, and $c$ are free parameters used to capture any potential large-scale velocity gradient.  We apply this model to the empirical velocity map displayed in the left-most panel of Figure~\ref{fig:TXS0206_w_grad}, and obtain the best-fit velocity gradient map as shown in the left-most panel of Figure \ref{fig:TXS0206_wo_grad}.  The gradient in the model is $\approx 3.7$ km/s/kpc. We estimate the uncertainty of this gradient by fitting 1000 velocity maps that are randomly generated based on the MCMC line fitting chain for each spaxel.  Due to the relatively large number of spaxels included in the analysis (i.e., over 2000 in the field of TXS0206$-$048), the velocity gradient based on this simple three-parameter model is well-determined.  We then subtract the best-fit 2D velocity gradient from the original velocity map and obtain the residual velocity map shown in Figure \ref{fig:TXS0206_wo_grad}.  

At first look, the best-fit unidirectional flow model does not completely capture the coherent flows displayed in Figure \ref{fig:TXS0206_w_grad}.  While it captures the apparent velocity shear along the east-west direction, the velocity gradient visible along the north-south direction remains.  This motivates a different approach to consider the presence of radial/tangential flows, which is discussed in \S\,\ref{sec:radial} below.  Here we proceed with the discussion using the residual map displayed in Figure \ref{fig:TXS0206_wo_grad}. 
We repeat the VSF measurements described in \S\,\ref{sec:VSF_measurement} and obtain both the $S^{\prime}_2$ and the $S^{\prime}_2$-$S^{\prime}_3$ ESS relation.  The results are shown in the right panels of Figure \ref{fig:TXS0206_wo_grad}.  

As expected, subtracting a large-scale velocity gradient has a larger impact on larger scales, and $S^{\prime}_2$, in general, becomes flatter compared to the results in Figure \ref{fig:TXS0206_w_grad} using the original velocity map.  Instead of continuing to rise to larger scales, $S^{\prime}_2$ appears to flatten at $r_{\rm proj}\approx 40$ kpc.  The $S^{\prime}_2$ vs.\ $S^{\prime}_3$ ESS still holds for the gradient removed velocity map. We estimate an intrinsic power-law slope of $\gamma_2=0.56^{+0.16}_{-0.17}$ for $S_2$ and $\gamma_3=0.78^{+0.28}_{-0.25}$ for $S_3$.  Note that the fitting range is now restricted to 8.3 kpc$<r_{\rm proj} <$40 kpc due to the flattening at 40 kpc, resulting in larger uncertainties in the best-fit slopes.  While the slope is flatter than what is obtained before removing the velocity gradient model, the two results are consistent to within the uncertainties.  Similarly, we also overplot the expected Kolmogorov $S^{\prime}_2$ with a slope of 2/3 after convolving with the PSF as the orange dashed curve in Figure \ref{fig:TXS0206_wo_grad}. It is clear that despite the data points exhibiting a flatter overall trend, the measurements still agree with the Kolmogorov slope over the scales probed. 


Based on the morphology of the nebulae (see Figure \ref{fig:nb}) and the velocity measurements, the nebulae in all four fields do not show signatures of well-established rotation disks.  We, therefore, do not consider a more elaborate disk model with additional parameters such as inclination and maximum rotation velocity.  

\subsection{Effects of radial and tangential motions}
\label{sec:radial}
Complementary to the simple, unidirectional coherent flows discussed above, here we investigate whether there exist significant differences between the VSFs constructed along the radial vs.\ tangential directions. This is a physically motivated scenario as gas outflows can manifest as coherent, radial motions while gas infalls are more likely to form large-scale tangential motions due to the conservation of angular momentum. For instance, if a nebula is mostly comprised of isotropic supergalactic winds, we would expect that the measured $S_2$ is driven by the power associated with radial motions with the best-fit slope $\gamma_2$ indicative of the acceleration of the wind.  
In addition, this test can also reveal anisotropy if the radial vs. tangential VSFs exhibit distinctive shapes.

\begin{figure}
	\includegraphics[width=\linewidth]{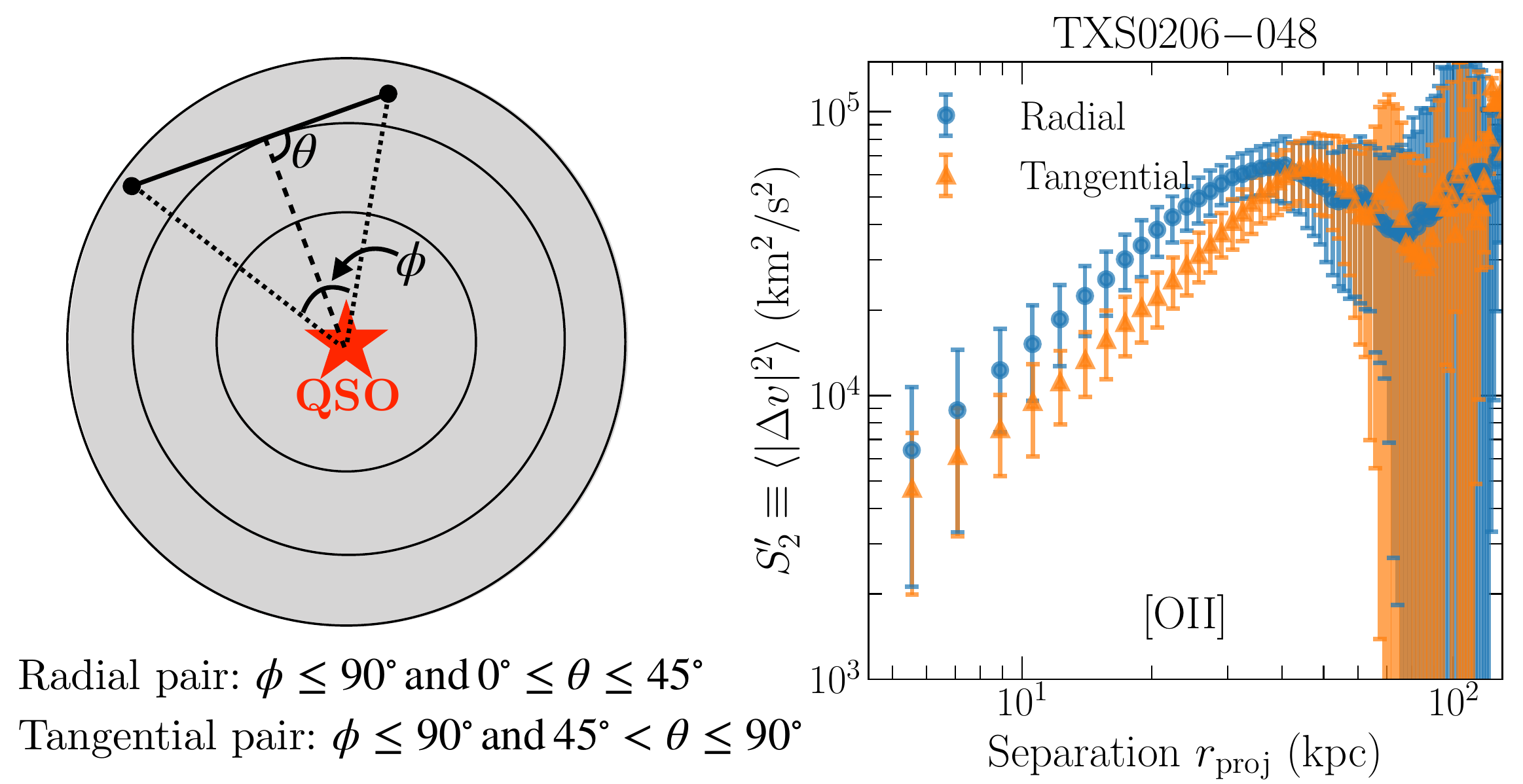}
    \caption{{\it Left:} Illustration of the radial vs.\ tangential pair classifications.  {\it Right:} The second-order VSFs $S^{\prime}_2$ measured using radial and tangential pairs, respectively, based on the velocity map presented in Figure \ref{fig:TXS0206_w_grad}. The shapes of $S^{\prime}_2$ for the radial and tangential pairs are consistent with each other, while the radial pairs exhibit a higher amplitude in the VSF. }
    \label{fig:tangential_vs_radial_cartoon}
\end{figure}

Using the velocity map presented in Figure \ref{fig:TXS0206_w_grad}, we classify the velocity pairs into two groups based on their spatial configuration with respect to the QSO location.  The classification criterion is illustrated in the left panel of Figure \ref{fig:tangential_vs_radial_cartoon}.  In this classification, we require both pixels in a pair to reside in the same quadrant of the nebula with the angle $\phi$ (see Figure \ref{fig:tangential_vs_radial_cartoon}) being equal to or smaller than $90^\circ$. Velocity pairs taken from pixels located in different quadrants of the nebula are not considered to avoid ambiguities between radial and tangential pairs.  We then calculate $\theta$, which is the angle between the vector that connects the two points in a pair and the vector that connects the pair mid-point to the QSO location, as shown in Figure \ref{fig:tangential_vs_radial_cartoon}. We assign any pairs with $0^\circ \le \theta \le 45^\circ$ ($45^\circ < \theta \le 90^\circ$) as radial (tangential) pairs, and repeat the VSF measurements using these two groups of pairs separately.  The results are shown in the right panel of Figure \ref{fig:tangential_vs_radial_cartoon}. 

The shapes of $S^{\prime}_2$ for the radial and tangential pairs are consistent with each other, while the radial pairs exhibit a slightly higher amplitude in the VSF.  This test demonstrates that the nebula gas undergoes 
dynamical processes with similar turbulence energy cascade characteristics
along the radial and tangential directions, and that both directions have comparable contributions to the signal in the total VSF presented in Figure \ref{fig:TXS0206_w_grad}.  Repeating this exercise with the gradient-removed velocity map leads to the same conclusion.

\begin{table*}
	\centering
	\caption{Summery of the power-law slopes of the VSFs constructed using [\ion{O}{II}] and [\ion{O}{III}] lines$^a$. }  
	\label{tab:slopes}
    \begin{threeparttable}
	\begin{tabular}{lcccccccc} 
		\hline
		 & \multicolumn{2}{c}{[\ion{O}{II}]} & \multicolumn{2}{c}{[\ion{O}{II}] grad.\ removed$^b$} & \multicolumn{2}{c}{[\ion{O}{III}]} & \multicolumn{2}{c}{[\ion{O}{III}] grad.\ removed$^b$} \\
		Field name &$\gamma_2$ &$\gamma_3$ &$\gamma_2$ &$\gamma_3$ &$\gamma_2$ &$\gamma_3$ &$\gamma_2$ &$\gamma_3$  \\
		\hline
		PKS0454$-$22 & $<0.78$ & $<1.15$ & $<0.66$ & $<0.99$
		             & $<0.67$ & $<0.94$ & $<1.45$ & $<2.3$\\[0.15cm]
		J0454$-$6116 & $<0.51$ & $<0.77$ & $<0.45$ & $<0.74$
		             & $<0.84$ & $<1.26$ & $<0.33$ & $<0.48$\\[0.15cm]
		J2135$-$5316 & $<0.50$ & $<0.76$ & $<0.65$ & $<1.02$
		             & $<1.23$ & $<1.81$ & $<1.12$ & $<1.75$\\[0.15cm]
		             
		TXS0206$-$048 & $0.72^{+0.12}_{-0.11}$ & $1.03^{+0.18}_{-0.16}$ & $0.56^{+0.16}_{-0.17}$ & $0.78^{+0.28}_{-0.25}$ &--&--&--&--  \\
		\hline
	\end{tabular}
	\begin{tablenotes}
	\footnotesize
    \item \textbf{Notes.}
    \item[{\it a}]  Constraints for the best-fit slopes listed here are based on the 1000 modified bootstrap samples (see \S\,\ref{sec:VSF_measurement}). These slopes are the intrinsic power-law slopes for $S_2$ and $S_3$, as our fitting procedure explicitly accounts for the smoothing effect in the measured $S^{\prime}_2$ and $S^{\prime}_3$ (see \S\,\ref{sec:VSF_of_txs0206}). For PKS0454$-$22, J0454$-$6116, and J2135$-$5316, we present 95\% upper limits for the slope under the assumption that the observed pair separations are within the inertial range.  If the available pair separations are close to injection scales, then no robust constraints can be obtained. For TXS0206$-$048, we list the median value as well as the 16$^{\rm th}$ and $84^{\rm th}$ quantiles as lower and upper limits. Note that, as discussed in \S\,\ref{sec:VSF_of_txs0206}, we only consider non-negative power-law slopes. 
    \item[{\it b}] Measurements obtained after removing a 2D velocity gradient (see \S\,\ref{sec:1Dgrad}).
    \end{tablenotes}
    \end{threeparttable}
\end{table*}

\subsection{The observed VSFs of PKS0454$-$22, J0454$-$6116 and J2135$-$5316}
\label{sec:VSF_other_3_fields}
For the remaining three fields, PKS0454$-$22, J0454$-$6116 and J2135$-$5316, both the \oii and \oiii lines are detected in the MUSE cubes.  We present the VSF measurements based on both lines, which are shown in Figures \ref{fig:PKS0454_o2}-\ref{fig:J2135_o3} in the Appendix. Constraints on the slopes of the VSFs are summarized in Table \ref{tab:slopes}. 

Compared with the results for TXS0206$-$048, the constraints on the slopes of the VSFs for these three QSO nebulae are weaker. The large uncertainties can be attributed to the limited dynamic range in spatial scale when comparing 
the spatial extent of the line-emitting nebulae with the size of the PSF in the data (see \S\,\ref{sec:VSF_bkg}). As listed in Table \ref{tab:emission_line_properties}, the [\ion{O}{II}] nebula included in the VSF measurements for TXS0206$-$048 is $\approx 2$--4 times larger than that of these three fields. The larger area leads to smaller uncertainties in the VSF measurements in each distance bin, and a larger dynamic range in distance separation, both contributing to a better-constrained VSF.  In contrast, a limited dynamic range in the pair separations for the 
remaining three nebulae inevitably pushes the VSF measurements closer to the injection scale, where we expect the VSF to flatter \citep[e.g.,][]{Benzi1993}.  If this is the case, then no robust constraints can be obtained for the VSF slopes in the inertial range. 


Similar to the result of TXS0206$-$048, removing a large-scale unidirectional velocity gradient from the velocity maps results in a flatter VSF. However, the measured slopes are consistent before and after the gradient removal, particularly with the large uncertainties for these fields. The VSFs calculated with radial vs.\ tangential pairs are also consistent in terms of the general shape and amplitude within each field, as shown in Figure \ref{fig:rad_tan_3field}.  Despite poorly constrained $S_2^{\prime}$, a strong correlation between $S_2^{\prime}$ and $S_3^{\prime}$ remains with $\gamma_2/\gamma_3\approx 0.7$. (see Figures \ref{fig:PKS0454_o2}--\ref{fig:J2135_o3}). 

\subsection{Effects of line-of-sight projections}
\label{sec:projection}

The availability of both \oii and \oiii emission signals for three of the QSO nebulae studied here also offers an opportunity to investigate the effect of line-of-sight  projection.  In particular, while velocity measurements of \oii and \oiii for PKS0454$-$22, J0454$-$6116 and J2135$-$5316 are mostly consistent with each other, there are regions with significantly different values between the two velocity maps, revealing not only that the emission signals are a blend of multiple components along the sightline but also that there exists a large variation  in the [\ion{O}{III}]/[\ion{O}{II}] line flux ratio between different components.  
Such variations indicate changing ionization conditions between different gas clumps that overlap along the line-of-sight and/or are unresolved along the plane of the sky (see \S\,\ref{sec:o32} for further discussion). 

Here we test how the measured VSFs change with different treatments of regions showing multi-component \oiii line profiles. Specifically, we compare three different scenarios where we assign to each multi-component spaxel (1) the velocity of the component with the largest line flux, (2) the velocity obtained by forcing a one-component fit, 
and (3) the flux-weighted mean velocity across all components.  We present the VSF comparison under these three scenarios in Figure \ref{fig:o3_multicomp_compare}. While the uncertainties are large, this data set indicates that using the velocity of the dominant component in flux for multi-component spaxels may lead to a flatter VSF with higher amplitudes on small scales.  Using the flux-weighted mean velocity and adopting a one-component fitting velocity results in similar VSFs, which motivates our decision to present the single-component fitting result in the VSF measurements.  

\section{Discussion}
\label{sec:discussion}

Of the four QSO nebulae studied in this work, we have shown that the VSFs of one QSO nebula, TXS0206$-$048, are in spectacular agreement with expectations from the Kolmogorov law.  The Kolmogorov model applies to isotropic, homogeneous, and incompressible flows. The observed agreement, therefore, implies that gas flows in the nebula are subsonic and that the turbulent energy is being transferred at a constant rate between different spatial scales. 
Given the expectation that the observed [\ion{O}{II}] emission traces cool gas of temperature $T\sim 10^4$ K with a sound speed of $c_s^{\rm cool}\approx 10$ km/s, the observed velocity difference of $\Delta\,v\gtrsim 100$ km/s on scales greater than 10 kpc would lead to a conclusion of supersonic motions within the cool gas.  On the other hand, the QSO is found to reside in a massive halo of $M_{\rm halo}\approx 5\times 10^{13}\msun$ (see \S\,\ref{sec:k41} below)
with an anticipated temperature of $T\sim 10^7$ K for the hot halo and a sound speed of $c_s^{\rm hot}\approx 300$ km/s.  If the [\ion{O}{II}]-emitting gas originates in cool clumps condensed out of the surrounding hot halo, then the observed VSFs capture the subsonic motions of individual clumps relative to the hot medium.
For the remaining three nebulae around PKS0454$-$22, J0454$-$6116, and J2135$-$5316, however, no robust constraints for the VSFs can be determined due to a limited dynamic range in seeing-limited data. 

In this section, we discuss the implications for the energy balance in the diffuse CGM in these QSO host nebulae. 
We first estimate the turbulence energy transfer rate, using TXS0206$-$048 as an example, and explore possible causes for the observed differences in the VSFs between the different QSO nebulae.  Finally, we review the limitations and caveats in the observations.

\subsection{Constant turbulent energy cascade in TXS0206$-$048}
\label{sec:k41}
For turbulent gas that follows the Kolmogorov law,  
the mean energy transfer rate per unit mass $\epsilon$ is expected to be constant within the inertial range and can be estimated following 
\begin{equation}
    \epsilon = \frac{5}{4}\left[\frac{|\langle \Delta v (r)^3\rangle|}{r}\right] \approx \frac{5}{4}\left[\frac{\langle |\Delta v (r)|^3\rangle}{r}\right].
\end{equation}
This is commonly referred to as the ``four-fifths law" in fully developed turbulence, and is an exact result derived from the Navier-Stokes equations \citep[][]{Kolmogorov1941, Frisch1995}. As stated in \cite{Benzi1993}, the relation $|\langle \Delta v (r)^3\rangle| \approx \langle |\Delta v (r)|^3\rangle$ is not obvious from first principles but has been experimentally verified.  Using the $S_3$ measurement for TXS0206$-$048, we obtain $\epsilon\approx0.2$ cm$^{2}$ s$^{-3}$. This energy transfer rate is comparable to the value measured with H$\alpha$ filaments in the Perseus cluster \citep[][]{Li2020}, as well as the 0.1--1 cm$^{2}$ s$^{-3}$ rate estimated for the Orion Nebula \citep[e.g.][]{Kaplan1970}. \cite{Rauch2001} reported a lower $\epsilon$ of $\sim 10^{-3}$ cm$^{2}$ s$^{-3}$ for \ion{C}{IV} absorbers at $z\approx 3$, suggesting that the CGM in high-redshift star-forming halos 
is less turbulent with a lower energy cascade rate \citep[see also][]{Rudie2019}.  However, due to the unspecified uncertainty in the VSF measurement in \cite{Rauch2001} and the different data set used, it is not conclusive whether the current discrepancy between our result and that of \cite{Rauch2001} is significant. 

Figures~\ref{fig:TXS0206_w_grad} and \ref{fig:TXS0206_wo_grad} show that the VSF of TXS0206$-$048 flattens at around $50$ kpc. Given that the statistical uncertainty in the VSF does not increase significantly until a scale of $\approx 80$ kpc, the turnover point at 50 kpc may be interpreted as 
the energy injection scale in this system.  In contrast, we do not detect signatures of the dissipation scale in all systems due to a fundamental limitation on the spatial resolution in seeing-limited observations. 

In addition, the estimated $\epsilon$ suggests that turbulent energy is subdominant in QSO host halos, as can be shown through the following calculations. 
The total  mass of the dark matter halo hosting TXS0206$-$048 is estimated to be $M_{\rm halo}\approx 5\times 10^{13}\msun$ \citep[][]{Johnson2022}.
Adopting a baryon fraction of $f_b\approx0.15$ \citep[][]{Planck2020}, we calculate a total baryonic mass within a radius of 50 kpc to be $\sim 3\times 10^{11}\msun$ for an NFW halo with a reasonable choice of halo concentration (i.e., between 4 and 10). This gives us a total turbulent energy transfer rate of $\dot{E}_{\rm turb}\sim 10^{44}$ erg s$^{-1}$, assuming that gas of all phases is perfectly coupled dynamically and that the turbulence cascade does not affect gas residing at distances much larger than $\approx 50$ kpc (i.e., the injection scale) from the halo center. Keeping these assumptions in mind, the turbulent energy that will eventually dissipate and heat up the gas in the CGM is $\sim 0.05\%$ of the bolometric luminosity of the QSO (see Table~\ref{tab:summary_table}), which is similar to the wind energy fraction observed in AGN outflows \citep[e.g.,][]{Fabian2012,Sun2017}. 
At the same time, 
this turbulent heating rate has the same order of magnitude as the \oii line luminosity. As we expect the 
gas to also cool through other forms of emission (e.g., \oiii, H$\alpha$ and \lya lines for the $\sim$10$^4$ K ionized phase), the turbulent heating rate is not sufficient to offset cooling of this gas in the vicinity of a luminous QSO. 

Finally, we note that in comparison to the remaining three QSO nebulae included in this study (see Table \ref{tab:summary_table}), TXS0206$-$048 occurs at the highest redshift QSO, $z_{\rm QSO}\approx 1.13$, and appears to reside in the highest-mass halo with a significant number of group members being super-$L_*$ galaxies and a large velocity dispersion \citep[][]{Johnson2022}.  The associated galactic environment may also play a significant role in driving the turbulence in the CGM, in addition to QSO outflows.

\subsection{Implications of the VSF slopes}

\label{sec:slope}

While the Kolmogorov theory has explicit predictions for the slopes of the VSF, a number of factors can impact the empirical measurements and should therefore be taken into account when interpreting the results.
As discussed in \S\,\ref{sec:VSF_bkg}, 
if the thickness of the nebulae along the line of sight is larger than the scales probed in the VSF, the projection effect will steepen the VSF.  If the nebulae identified around these four QSOs are more sheet-like than spherical, then we would expect the intrinsic slope to be flatter than measured.  Of the four nebulae studied here, J2135$-$5316 exhibits an elongated morphology and is most likely affected by such projection effect.

One possible explanation for the flat slopes is the presence of dynamically important magnetic fields, where the kinetic energy cascade is suppressed due to magnetic tensions \citep[e.g.,][]{Boldyrev2006, Brandenburg2013,Grete2021,Mohapatra2022}.  Another interesting scenario for flattened VSFs is one where energy injections happen at multiple different length scales, instead of one scale that defines the canonical upper limit of the turbulent inertial range. When combining multiple kinetic energy power spectra with different injection scales, the resulting VSF reflects the superposition of the different components, leading to a flatter slope due to elevated power at scales smaller than the largest injection scale of the system \citep[e.g.,][]{ZuHone2016}. This scenario is consistent with a diverse range of dynamical processes expected to be present in the CGM of a QSO halo, such as gas outflows, mergers, AGN-inflated bubbles, and relativistic jets \cite[e.g.,][]{Fabian2012}.  While the detailed mechanisms through which these processes transfer kinetic energy to the gas are poorly understood at the current moment, it is likely that different processes have different characteristic scales for energy injection. Irrespective of what the detailed mechanisms are, if the pair separations are indeed closer to the injection
scale, then no conclusive constraints
can be obtained for the VSF slopes in the inertial range.

Alternatively, the range of VSF slopes across the four fields could also be suggestive of a time-dependent evolution of these nebulae. As the energy injection from QSO outflows is expected to be episodic, turbulent energy may be dissipated during the off cycle, 
leading to a flat VSF. 
For virialised systems with a complete absence of turbulence, \cite{Melnick2021} indeed obtains flat VSFs based on N-body simulations. 
Taking TXS0206$-$048 as a reference, turbulent energy on scales of $\sim 50$ kpc is expected to be dissipated on a time scale of $\langle |\Delta v|^2\rangle/\epsilon\sim 100$ Myr, and the time scale will be shorter for smaller spatial scales.  Under this scenario, the observed flatter VSFs in the three lower redshift QSO nebulae suggest that the most recent episode of significant energy injection 
occurred more than $\sim 100$ Myr ago. Because such time scale exceeds the typical QSO lifetime of $\sim 0.1-10$ Myr \citep[e.g.][]{Schawinski2015,Sun2017,Shen2021}, this would make the radiative feedback during the luminous phase of an AGN an unlikely source for driving the observed turbulence.

At the same time, recall that the four QSOs reside in a diverse range of galactic environments, with TXS0206$-$048 in a rich dynamic galaxy group while J2135$-$5316 in a relatively isolated environment with only two neighboring galaxies found (see Table \ref{tab:summary_table}).  If galaxy/satellite interactions are a main driver of the turbulent CGM, then a flat VSF found for J2135$-$5316 may be attributed to the quiescent state of its galactic environment.

\subsection{Extended self-similarity scaling slopes}
\label{sec:ESS_discussion}
In addition to the slopes $\gamma_p$ of individual VSFs, the ESS scaling slopes between different orders can also shed light on the dynamic state of the gas. 
We have measured the slopes of VSFs of each nebulae for up to $p=6$. 
As mentioned in \S\,\ref{sec:VSF_of_txs0206}, with the current data set, VSFs for $p>6$ become too noisy to deliver meaningful constraints.  The results are presented in Figure \ref{fig:ESS_slopes}, along with 
theoretical expectations presented in \cite{SheLeveque1994} and \cite{Boldyrev2002}.  These models account for Kolmogorov turbulence with the intermittency correction and supersonic magnetohydrodynamic turbulence, respectively. The simulation results for transonic and supersonic hydrodynamic turbulence presented in \cite{PanScannapieco2011} are also included in Figure \ref{fig:ESS_slopes} for comparison. 

It is clear that the strongest discriminating power between these different scenarios lies in the higher-order VSFs with $p\ge 4$. 
Due to large uncertainties in our measurements particularly for higher orders, we can only rule out the scenario for supersonic hydrodynamic turbulence with a Mach number of 6.1 from \cite{PanScannapieco2011}.  While the measurements appear to support the presence of subsonic turbulence in all four quasar nebulae, we note that the simple $p/3$ scaling relative to $S_3$ is also expected from a simple dimensional inference. Consequently, in the absence of direct measurements of $S_2$, 
the relative scaling between different orders alone does not provide conclusive evidence for whether or not the gas follows subsonic turbulence.

\subsection{[\ion{O}{II}] and [\ion{O}{III}] surface brightness profiles}
\label{sec:o32}

\begin{figure*}
	\includegraphics[width=\linewidth]{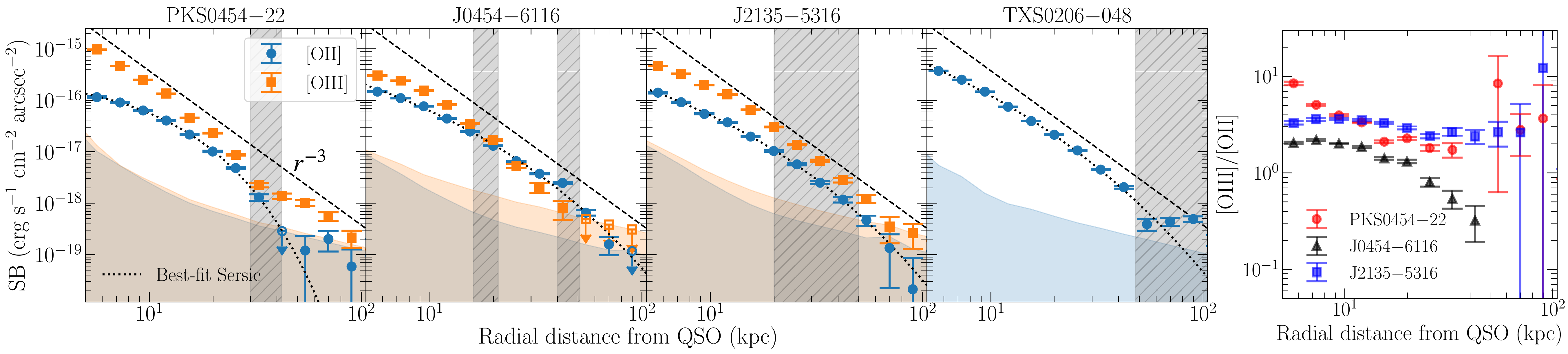}
    \caption{Surface brightness profiles of the four fields, and the corresponding [\ion{O}{III}]/[\ion{O}{II}] line flux ratios for the three lower redshift fields. The surface brightness profiles are circularly averaged within annuli at different distances from the QSOs.  For the first three fields, vertical shaded regions indicate radii with relatively strong flux contributions from areas that are eliminated from the VSF measurements (see \S\,\ref{sec:VSF_measurement}). For TXS0206$-$048, the vertical shaded region indicates the radii of the stream-like structures away from the main nebula (see Fig.~\ref{fig:nb}). Blue and orange shaded regions at the bottom of each panel show the $2\sigma$ limit of the [\ion{O}{II}] and [\ion{O}{III}] surface brightness level as a function of radius. Dotted curves show the best-fit S\'{e}rsic profiles for the [\ion{O}{II}] line, with half-light radius $R_e\approx[10, 10, 9, 6]$ kpc and S\'{e}rsic index $n\approx[1.1, 2.6, 2.5, 3.0]$ for the four fields from left to right. For the [\ion{O}{III}] profiles, however, we cannot find a good fit with S\'{e}rsic, exponential, or cored isothermal profiles. Instead, we overplot a power-law with $-3$ slope in dashed lines for comparison, as it provides a good match to the [\ion{O}{III}] profile in PKS0454$-$22, as well as the [\ion{O}{II}] and [\ion{O}{III}] profiles outside of the core ($\approx 10$ kpc) region in other fields. In the right-most panel, [\ion{O}{III}]/[\ion{O}{II}] line flux ratios are shown for the three lower redshift fields, and data points with only upper limits on [\ion{O}{II}] or [\ion{O}{III}] are not included. }
    \label{fig:sb_profile}
\end{figure*}

As mentioned in \S\,\ref{sec:projection}, examinations of the observed [\ion{O}{III}]/[\ion{O}{II}] line flux ratio across the nebulae have revealed intrinsic differences in the gas traced by the [\ion{O}{II}] and [\ion{O}{III}] emission features.  Here we investigate the circularly-averaged radial surface brightness profiles for [\ion{O}{II}] and [\ion{O}{III}] lines, as shown in Figure~\ref{fig:sb_profile}, in order to gain insights into the difference in the spatial distribution of the gas probed by different emission features. The observed one-dimensional surface brightness profile also facilitates a direct comparison of the gas properties across the four fields and with high-redshift quasars.  Similar to the practice in \cite{Borisova2016}, to obtain a more robust uncertainty estimate for the surface brightness level, we use narrow-band images collapsed over a fixed range of wavelength slices across the whole field, instead of the optimally subtracted images shown in Figure~\ref{fig:nb}.  The wavelength range used for the narrow-band images here is decided based on the largest range along the wavelength dimension in the corresponding 3D masks (see \S\,\ref{sec:nb_3dmask}). We also manually remove additional residuals in the narrow-band images that are not associated with the nebulae.  This step is necessary because taking the circularly averaged value within each annulus could pick out faint spurious signals, especially in the noise-dominated regions. For the areas that are filtered out in the VSF measurement step (see \S\,\ref{sec:VSF_measurement}), we indicate their corresponding radii with vertical shaded regions in Figure~\ref{fig:sb_profile} to guide the visual comparison. For TXS0206$-$048, the vertical shaded region denotes where the stream structures away from the main nebula contribute significantly to the averaged surface brightness level. 

We find a good fit for the [\ion{O}{II}] emission with S\'{e}rsic profiles \citep[][]{sersic1968}, with the best-fit half-light radius $R_e\approx[10, 10, 9, 6]$ kpc and the best-fit S\'{e}rsic index $n\approx[1.1, 2.6, 2.5, 3.0]$, for the four fields with increasing redshifts. The best-fit models for the [\ion{O}{II}] profiles are shown in dotted curves in Figure~\ref{fig:sb_profile}. For the [\ion{O}{III}] profiles, however, we cannot find a good fit with S\'{e}rsic, exponential, or cored isothermal profiles. Instead, a single power-law with a slope of $\approx -3$ can provide a good match to the [\ion{O}{III}] profiles, except for the flat core regions (approximately inner 10 kpc) of J0454$-$6116 and J2135$-$5316.  We therefore simply overplot this power-law with a slope of $-3$ in Figure~\ref{fig:sb_profile} for comparison.  This slope also roughly matches the slopes of the [\ion{O}{II}] profiles outside of the core region.

Note that the [\ion{O}{II}] and [\ion{O}{III}] surface brightness profiles in the optical nebulae here are much steeper than the spatial profiles observed in extended \lya nebulae around $z\approx3$ QSOs, which have characteristic power-law slopes of $\approx -2$ \citep[e.g.,][]{Steidel2011, Borisova2016, Battaia2019}.  This contrast in slope between optical nebulae and \lya nebulae can be explained by the resonant nature of \lya photons, resulting in more extended \lya emission with shallower spatial profiles compared with the continuum and non-resonant line emission \citep[e.g.,][]{Steidel2011,Wisotzki2016,Patricio2016,Leclercq2017,Chen2021}. 

In the right-most panel of Figure~\ref{fig:sb_profile}, we show the [\ion{O}{III}]/[\ion{O}{II}] line flux ratios as a function of radial distance from the QSOs for the three lower redshift fields. Here we see the manifestation of the extreme ionization condition in the vicinity of these bright QSOs, with the line ratios far exceeding the nominal values of [\ion{O}{III}]/[\ion{O}{II}]$<1$ for typical star-forming and even AGN regions \citep[e.g.,][]{Kewley2001,Kauffmann2003}. Particularly for PKS0454$-$22, the [\ion{O}{III}]/[\ion{O}{II}] ratio is significantly enhanced in the central 10 kpc, reaching a value of $\approx10$ at its peak. Interestingly, among the three lower redshift fields, the [\ion{O}{III}]/[\ion{O}{II}] line flux ratios as a function of spatial distance from the QSO exhibit different profiles. This difference confirms that significant variations in the underlying physical conditions, such as density, metallicity, and local ionizing radiation intensities, are present both within individual nebulae and between fields.  However, quantifying the impact on the VSF measurements will require higher signal-to-noise data.

\subsection{On the detection rate of QSO nebulae and its implications for turbulence studies}  

While the four QSO nebulae studied here exhibit a range of VSF slopes, a remaining question is how the results from this sample bear on quasar host halos as a  whole.  A fundamental limitation of the VSF measurements is the detectability of the diffuse gas, which is a combined result of instrument sensitivity and the physical conditions of the gas.  
Using the CUBS sample of 15 UV-bright QSOs \citep[][]{CUBS1}, the detection rate of extended optical QSO nebulae (i.e., $\gtrsim30$ kpc above the surface brightness level of $\sim 10^{-18}$ $\sbunit$) at $z\lesssim 1.5$ is $\approx 25$\%\footnote{Two of the four fields are presented in this study (i.e., J0454$-$6116 and J2135$-$5316), and two are not considered here due to their smaller sizes.}. 
While a more comprehensive search of the MUSE data archive is needed to better quantify the detection rate of extended optical nebulae around QSOs, the $\approx 25\%$ detection rate from the CUBS program most likely represents a conservative lower limit to the rate of incidence of extended nebulae around luminous QSOs.  
It remains to be determined as to whether deeper observations will both increase the detection rate of extended nebulae and uncover missing light at larger distances and lower flux levels.

With the current small sample size, no clear correlation is found between global QSO properties (e.g., luminosity, radio-loudness, number of group member galaxies) and the presence (or lack thereof) of extended optical nebulae. The current detection rate of extended nebulae around low-redshfit QSOs is in stark contrast with the 100\% detection rate of extended \lya nebulae around QSOs at $z\approx3$, and could be a result of the possible redshift evolution of the cool ($\sim 10^4$ K) gas content at different epochs \citep[e.g.,][]{Borisova2016,Battaia2019}.  However, a statistical sample of sources observed both in \lya and non-resonant lines over cosmic time has yet to be established for a rigorous investigation of the apparent discrepancy in the incidence of extended nebulae between QSOs at low and high redshifts. 

Meanwhile, evidence suggests that these nebulae could have a diverse range of physical origins.  In addition to different [\ion{O}{III}]/[\ion{O}{II}] ratios (see \S\,\ref{sec:o32} and Figure~\ref{fig:sb_profile}), the morphology and the kinematics of nebulae also provide important clues.  For example, the ``Makani" nebula exhibits morpho-kinematics that strongly suggests supergalactic winds being a predominant driver of the line-emitting region \citep[][]{Rupke2019}.  For PKS0454$-$22, 
the morpho-kinematics of the nebula and the continuum sources in the immediate vicinity of the QSO have led \cite{Helton2021} to argue that the extended nebula mostly consists of striped ISM through interactions between gas-rich galaxies. A similar case is made for the nebula in PKS0405$-$123 in \cite{Johnson2018} and TXS0206$-$048 \citep[][]{Johnson2022}. However, this scenario of ISM stripping does not seem to be plausible for J0454$-$6116 and J2135$-$5316 studied here. For J2135$-$5316, only two group member galaxies (both far away from the location of the nebula) are found in the QSO field with the current data set (see Table~\ref{tab:summary_table}). Similarly for the J0454$-$6116 field, although two continuum sources are found near the QSO, their velocities are inconsistent with the rest of the nebula and are likely not in a coherent dynamical system with the line-emitting gas (see \S\,\ref{sec:VSF_measurement}).  No additional companion continuum sources are found in this field that overlap with the nebula footprint.  Interestingly, both J0454$-$6116 and J2135$-$5316 exhibit a relatively flat VSF.  Future studies based on a larger sample are needed to investigate the respective roles of supergalactic winds and galactic environments in driving the turbulence of the CGM.


\subsection{Limitations and caveats}

A primary limitation of the current study is the relatively small dynamic range of length scales available for the VSF measurements. Specifically, the smallest scale accessible is limited by the FWHM of the effective PSF, which is a combination of the seeing disk in ground-based observations and the smoothing kernel applied to the final combined cubes to increase the SNR in the data (see \S\,\ref{sec:data}).  The largest scale is dictated by the size of the nebulae over which robust line signals can be measured.  When measuring the slope of the VSF, the dynamic range is further restricted to where a single power-law can provide an adequate description 
(see \S\,\ref{sec:VSF_of_txs0206}).  Uncertainties in the VSF 
have also led to ambiguities in drawing conclusions on the dynamical properties of the gas.  Even for TXS0206$-$048, the range of distance scales probed is less than a decade.  One possible way to increase the dynamic range is to target nebulae at lower redshifts.  For example, at the same physical size, a nebula at $z\approx0.1$ will be approximately $5\times$ larger in the apparent angular size than those at $z\approx 1$, enabling VSF measurements on smaller scales for a fixed seeing disk size. Alternatively, to improve the measurements for nebulae at high redshifts, it is necessary to reduce the size of effective PSF in the data. The infrared spectrograph, NIRSpec, onboard the {\it James Webb Space Telescope} ({\it JWST}) will deliver a PSF 10 times smaller than the natural seeing disk on the ground.  Using the 
upgraded adaptive optics assisted Narrow-Field-Mode provided by MUSE will also offer additional spatial resolving power 
for probing the energy power spectrum on scales as small as $\sim\,1$ kpc, but will require long exposures to reach sufficient SNR. 

An improved spatial resolution also helps to reduce systematic uncertainties in the two-dimensional VSF measurements due to blending of distinct structures between adjacent sightlines.
In analysing the \oii emission lines in all four fields, there is clear  evidence for large density variations across individual nebulae based on the doublet 
ratio.  
If a large density contrast exists on scales smaller than the spatial resolution kernel, then blending would also suppress the power on small scales.  Despite these caveats, it is interesting to see that the VSF measurements of three out of four nebulae in this study display a non-zero slope, indicating a clear scale-dependent power in the velocity structures.


\section{Conclusion}
\label{sec:conclusion}
In this study, we present measurements of the velocity structure functions for four optical nebulae detected in the vicinities of UV-luminous QSOs at $z\approx 0.5$-1.1. 
Using wide-field integral field spectroscopic data obtained from VLT/MUSE, we measure spatially-resolved kinematics using the \oii and \oiii emission lines, and construct VSFs based on these velocity maps. Out of the four field, one field (i.e., TXS0206$-$048 with the largest nebula area and the highest SNR in the VSF measurement) exhibits a second-order VSF consistent with Kolmogorov, suggesting that the gas flows are isotropic and subsonic.  We estimate a turbulent energy cascade rate of $\epsilon\approx0.2$ cm$^{2}$ s$^{-3}$.  The remaining three fields show a range of VSF slopes, while all being flatter than the Kolmogorov slope. Possible interpretations of the range of VSF slopes across the four fields include the presence of a dynamically important magnetic field, turbulent energy injection at multiple spatial scales, a time-dependent evolution of the turbulent motions in the nebulae, and the impact from the diverse range of galactic environment associated with different fields. Alternatively, the apparent flat slopes in the VSFs may simply be due to a lack of dynamic range in the pair separations for probing the inertial range, which can be directly tested with high spatial resolution IFS data to extend the VSF measurements to smaller scales.

We develop the methodology to explicitly account for the spatial correlation in the data due to atmospheric seeing and smoothing.  We also investigate possible contributions to the VSF measurements from a unidirectional velocity gradient, and large-scale radial or tangential rotational flows. These methods can be applied in future studies to obtain more robust VSF measurements. Our results improve upon traditional line width studies for inferring turbulent velocity fields in diffuse gas and provide a robust description of the energy power spectrum of the velocity field.  The findings of this study can be compared with high-resolution numerical simulations to further our understanding of the driving and development of turbulence in the CGM, and the impact of quasar feedback on the CGM dynamics specifically in the case of quasar nebulae. 

\section*{Acknowledgements}
MCC and HWC are grateful to Fausto Cattaneo for numerous enlightening  discussions on turbulence and fluid dynamics that helped guide the analysis presented in this paper.  We also thank the referee, Evan Scannapieco, for constructive comments that helped improve this paper. We thank Irina Zhuravleva, Yuan Li, Valeria Olivares, Yuanyuan Su, Judit Prat, Lucas Secco, Andrey Kravtsov, and Nick Gnedin for helpful discussions on various observational and theoretical issues throughout this work.  We also thank Yuan Li for sharing her VSF calculation code in the early stage of this analysis which helped jump start the exploration of our velocity cubes.  HWC and MCC acknowledge partial support from HST-GO-15163.001A and NSF AST-1715692 grants. ZQ acknowledges partial support from HST-GO-15163.001A and NASA ADAP grant 80NSSC22K0481. EB acknowledges support by NASA under award number 80GSFC21M0002. SC gratefully acknowledges support from the European Research Council (ERC) under the European Union’s Horizon 2020 research and innovation programme grant agreement No 864361. This research has made use of the services of the ESO Science Archive Facility and the Astrophysics Data Service (ADS)\footnote{\url{https://ui.adsabs.harvard.edu/classic-form}}. The analysis in this work was greatly facilitated by the following \texttt{python} packages:  \texttt{Numpy} \citep{Numpy}, \texttt{Scipy} \citep{Scipy}, \texttt{Astropy} \citep{astropy:2013,astropy:2018}, \texttt{Matplotlib} \citep{Matplotlib}, and \texttt{MPDAF} \citep{MPDAF}.  

\section*{Data Availability}
The data used in this article are available for download through the the ESO Science Archive Facility.



\bibliographystyle{mnras}
\bibliography{main} 




\appendix

\section{Some extra material}
Here we present the VSFs measurements for PKS0454$-$22, J0454$-$6116 and J2135$-$5316.  The results are discussed in \S\,\ref{sec:VSF_other_3_fields} and \S\,\ref{sec:discussion}.

\begin{figure*}
	\includegraphics[width=0.9\linewidth]{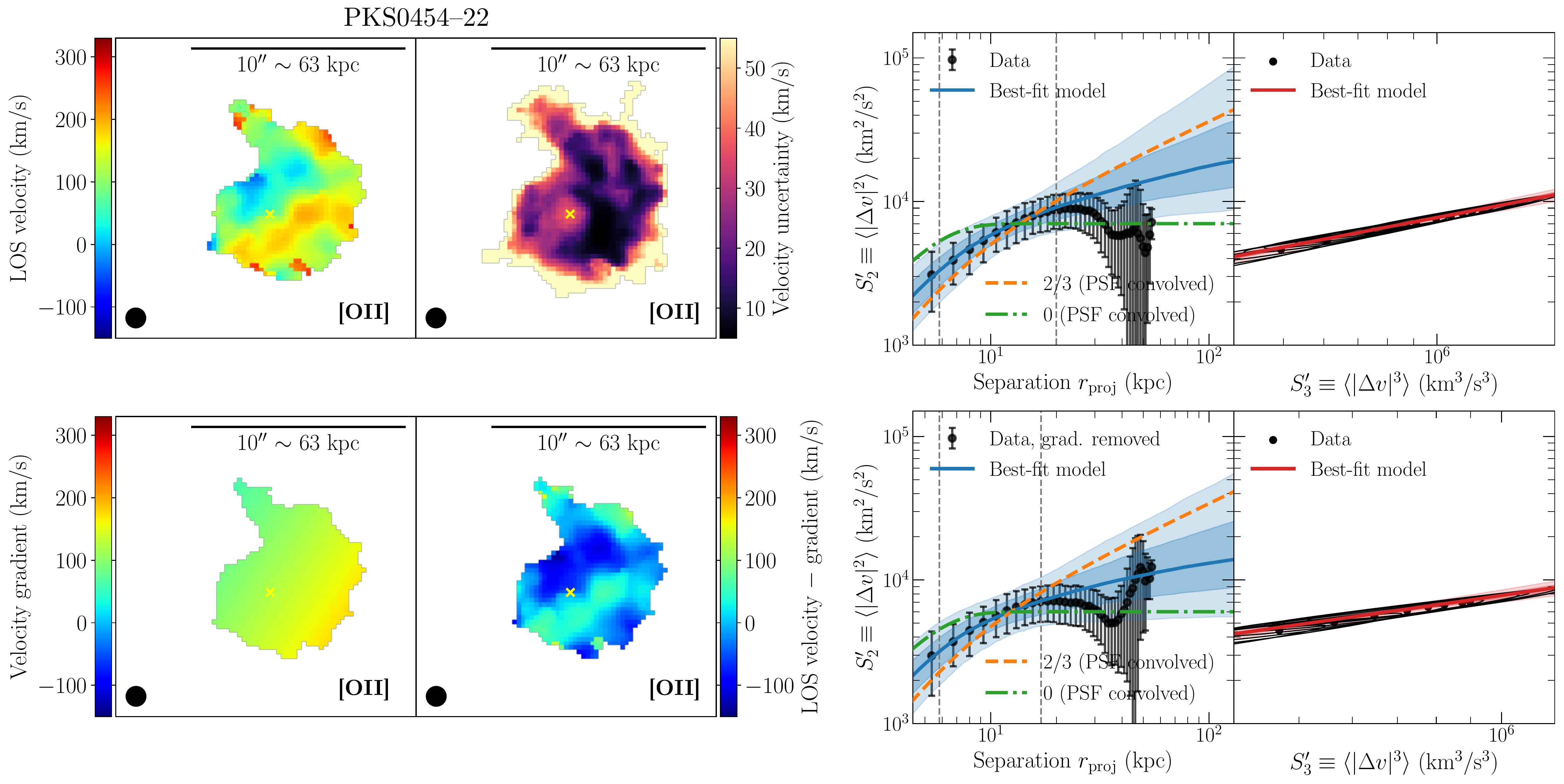}
    \caption{Same as Figs.~\ref{fig:TXS0206_w_grad} and \ref{fig:TXS0206_wo_grad} in the main text, but for the field of PKS0454$-$22 using the \oii emission line. Here a flat VSF (with a slope of 0) is also shown by the dotted-dash green line for comparison.}
    \label{fig:PKS0454_o2}
\end{figure*}

\begin{figure*}
	\includegraphics[width=0.9\linewidth]{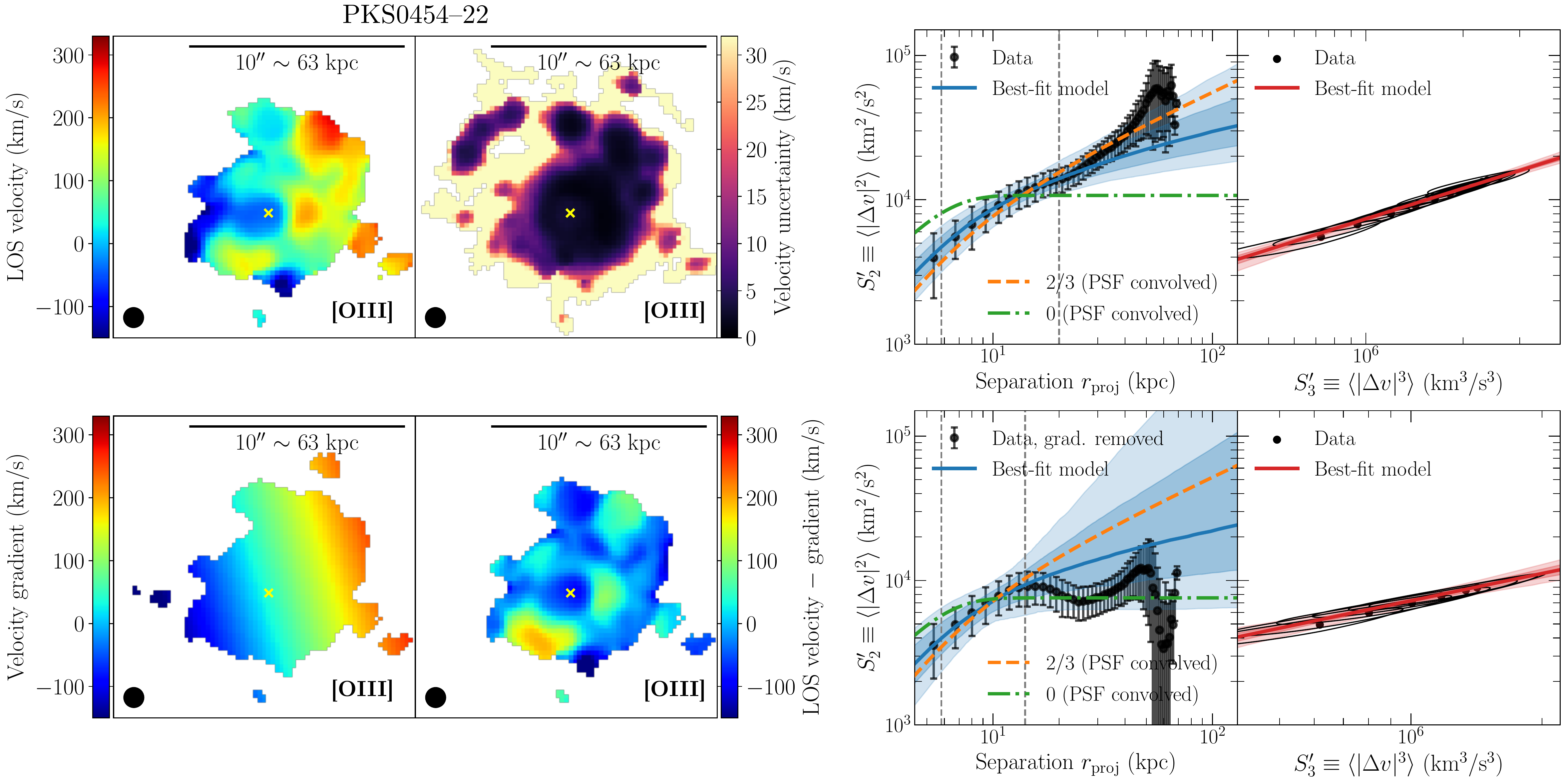}
    \caption{Same as Figs.~\ref{fig:TXS0206_w_grad} and \ref{fig:TXS0206_wo_grad} in the main text, but for the field of PKS0454$-$22 using the \oiii emission line. Here a flat VSF (with a slope of 0) is also shown by the dotted-dash green line for comparison.}
    \label{fig:PKS0454_o3}
\end{figure*}

\begin{figure*}
	\includegraphics[width=0.9\linewidth]{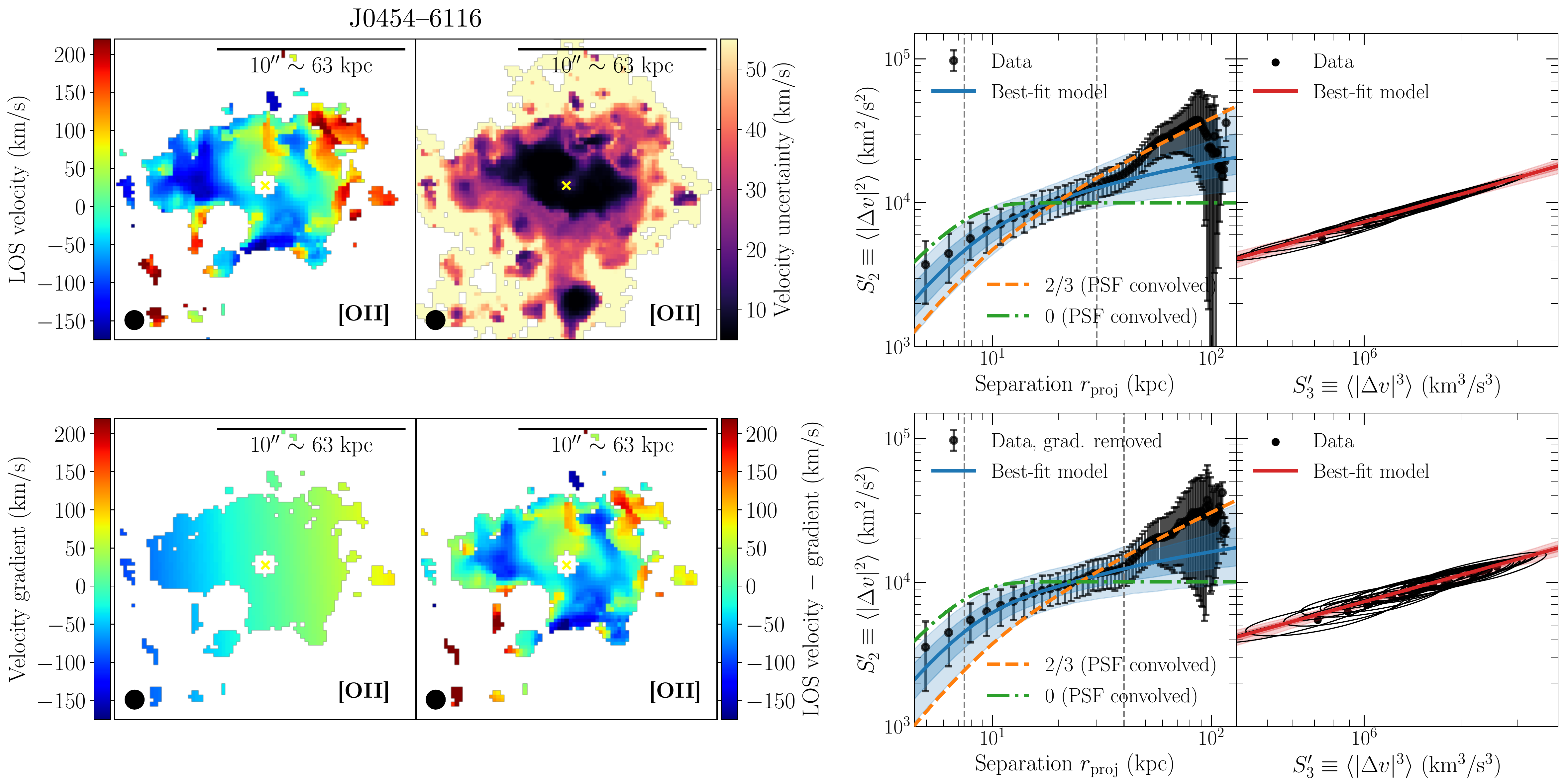}
    \caption{Same as Figs.~\ref{fig:TXS0206_w_grad} and \ref{fig:TXS0206_wo_grad} in the main text, but for the field of J0454$-$6116 using the \oii emission line. Here a flat VSF (with a slope of 0) is also shown by the dotted-dash green line for comparison.}
    \label{fig:J0454_o2}
\end{figure*}

\begin{figure*}
	\includegraphics[width=0.9\linewidth]{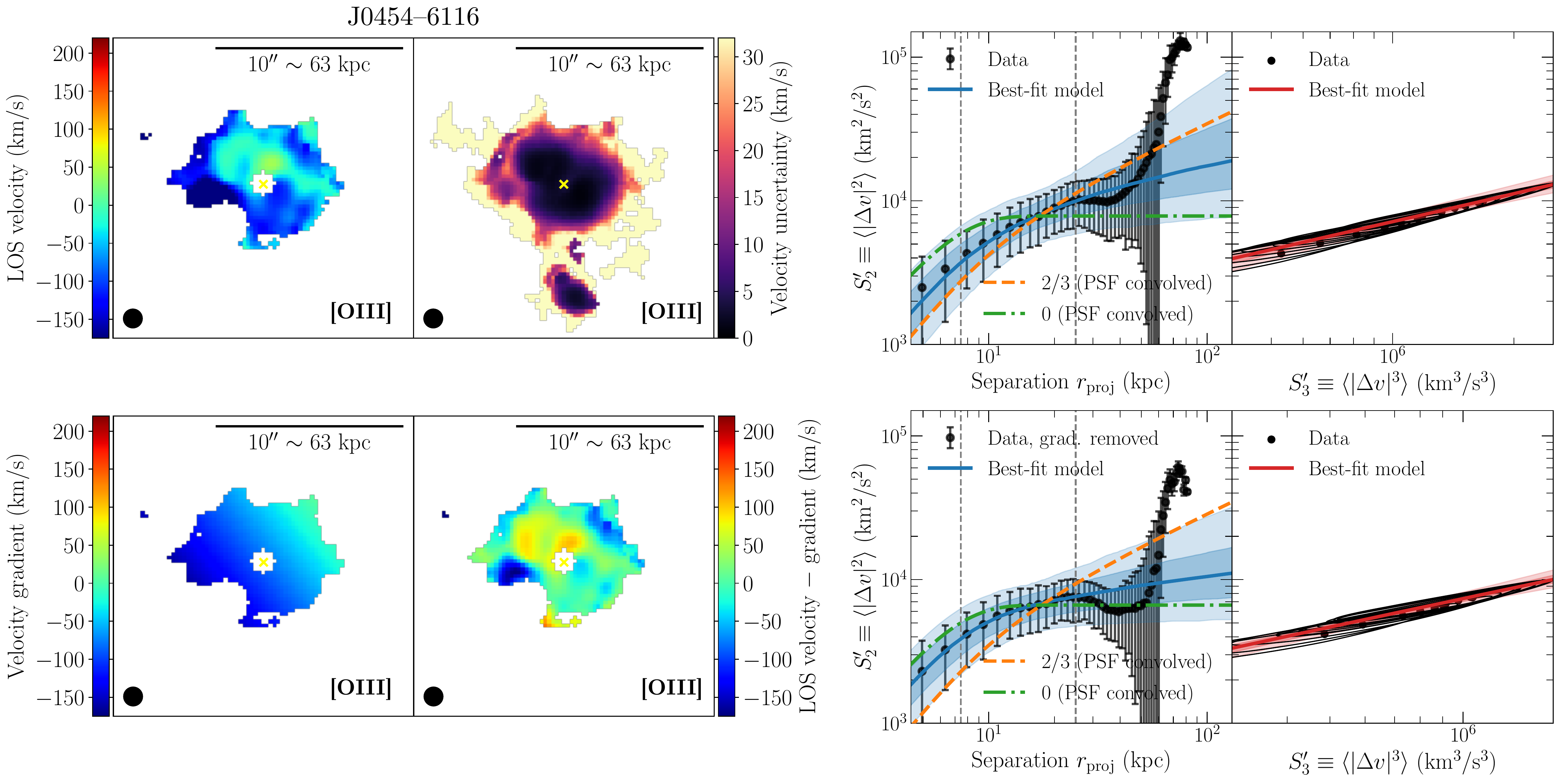}
    \caption{Same as Figs.~\ref{fig:TXS0206_w_grad} and \ref{fig:TXS0206_wo_grad} in the main text, but for the field of J0454$-$6116 using the \oiii emission line. Here a flat VSF (with a slope of 0) is also shown by the dotted-dash green line for comparison.}
    \label{fig:J0454_o3}
\end{figure*}

\begin{figure*}
	\includegraphics[width=0.9\linewidth]{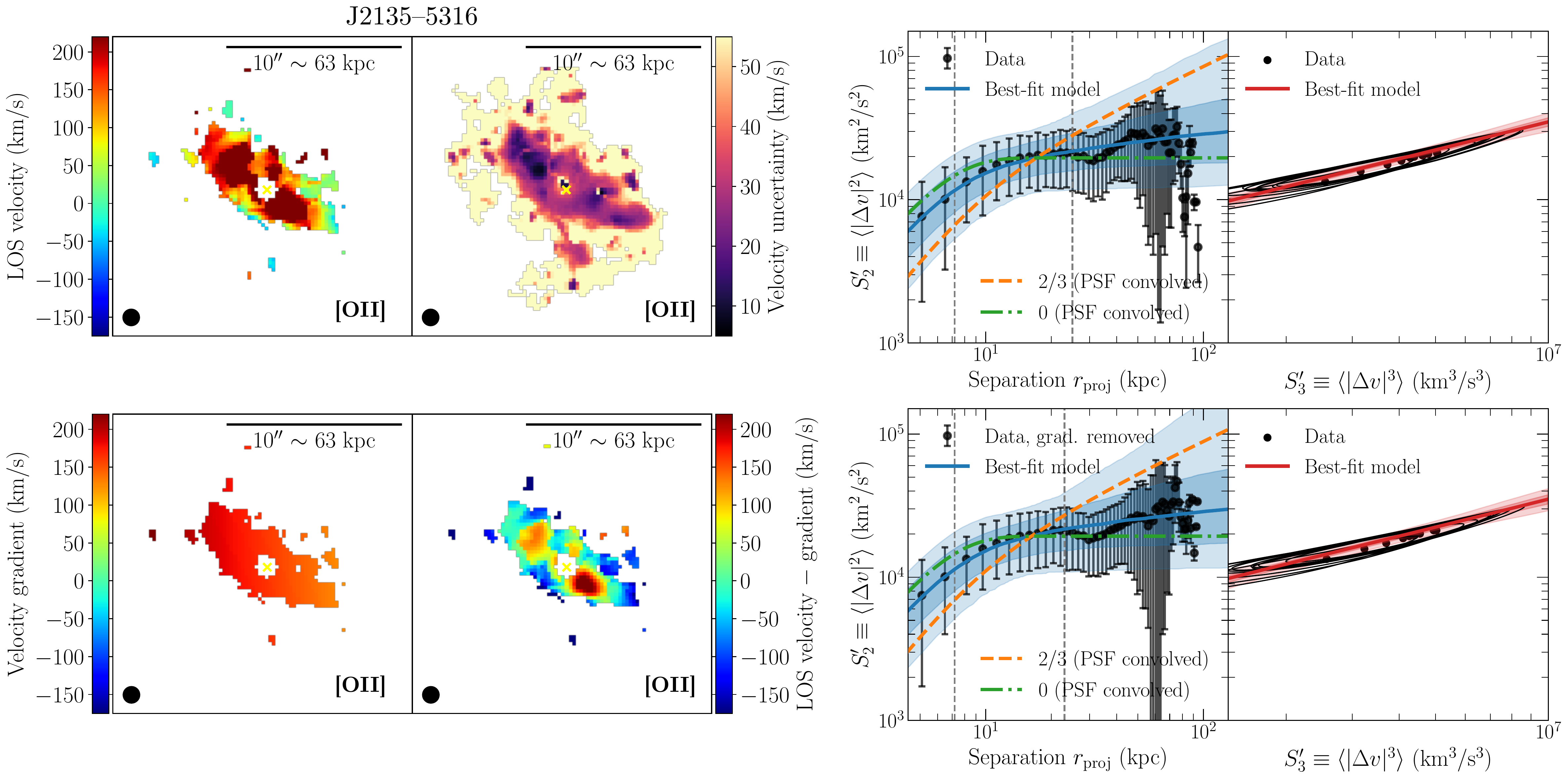}
    \caption{Same as Figs.~\ref{fig:TXS0206_w_grad} and \ref{fig:TXS0206_wo_grad} in the main text, but for the field of J2135$-$5316 using the \oii emission line. Here a flat VSF (with a slope of 0) is also shown by the dotted-dash green line for comparison.}
    \label{fig:J2135_o2}
\end{figure*}

\begin{figure*}
	\includegraphics[width=0.9\linewidth]{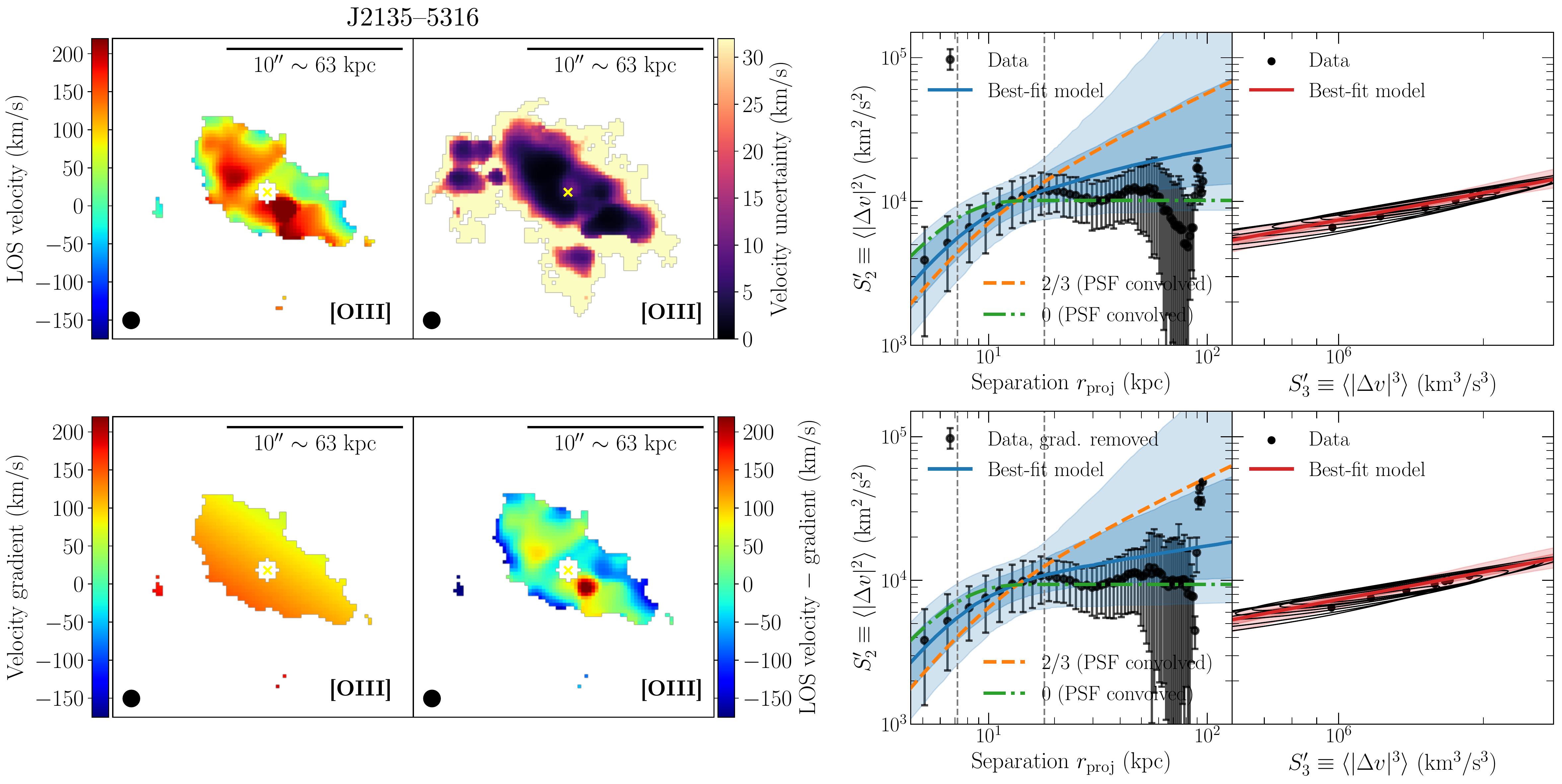}
    \caption{Same as Figs.~\ref{fig:TXS0206_w_grad} and \ref{fig:TXS0206_wo_grad} in the main text, but for the field of J2135$-$5316 using the \oiii emission line. Here a flat VSF (with a slope of 0) is also shown by the dotted-dash green line for comparison.}
    \label{fig:J2135_o3}
\end{figure*}




\begin{figure*}
    \centering
    \includegraphics[width=0.7\linewidth]{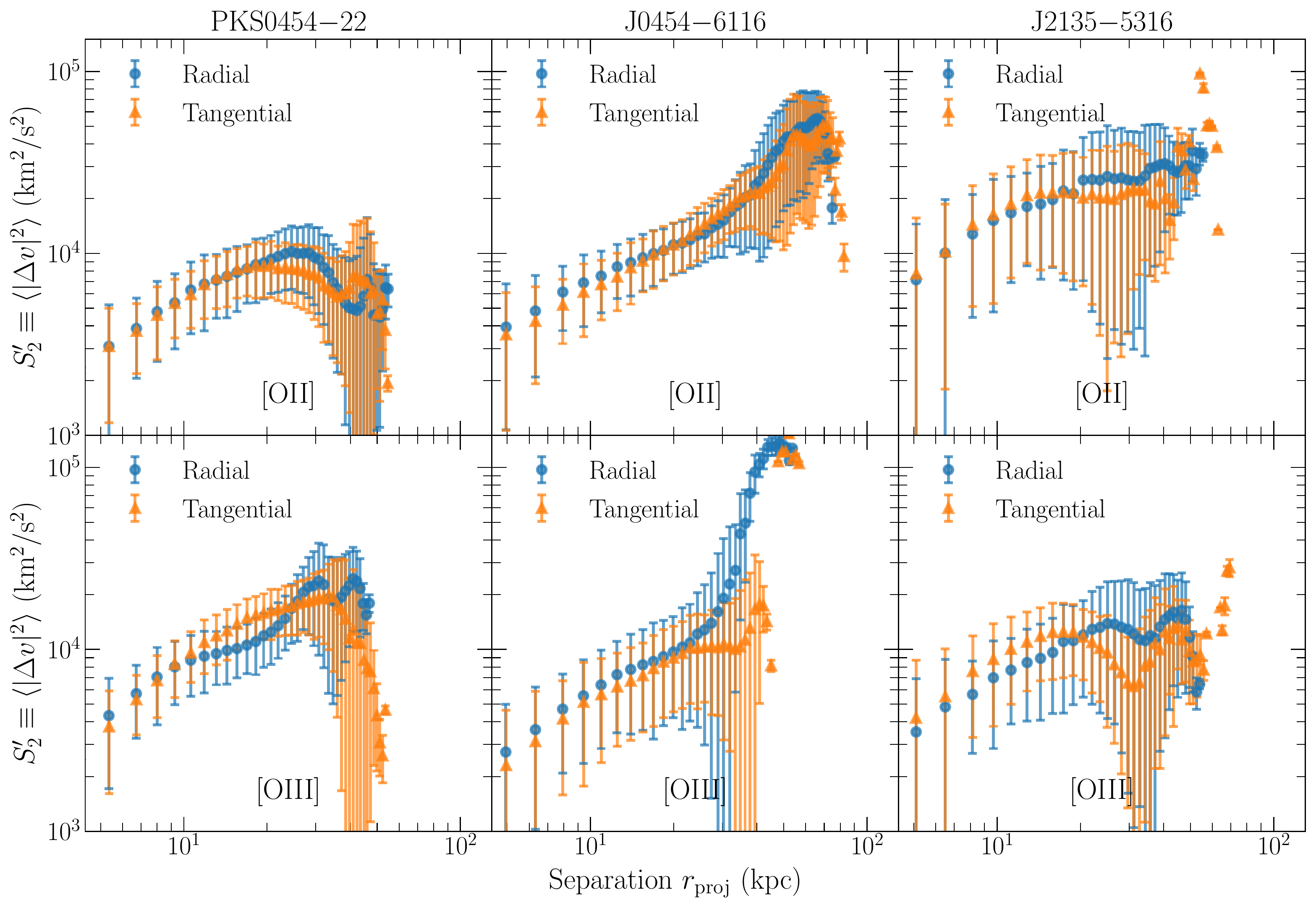}
    \caption{Same as the right panel of Fig~\ref{fig:tangential_vs_radial_cartoon}, but for the fields of PKS0454$-$22, J0454$-$6116, and J2135$-$5316, including results based on both \oii and \oiii emission lines. }
    \label{fig:rad_tan_3field}
\end{figure*}

\begin{figure*}
    \centering
    \includegraphics[width=0.7\linewidth]{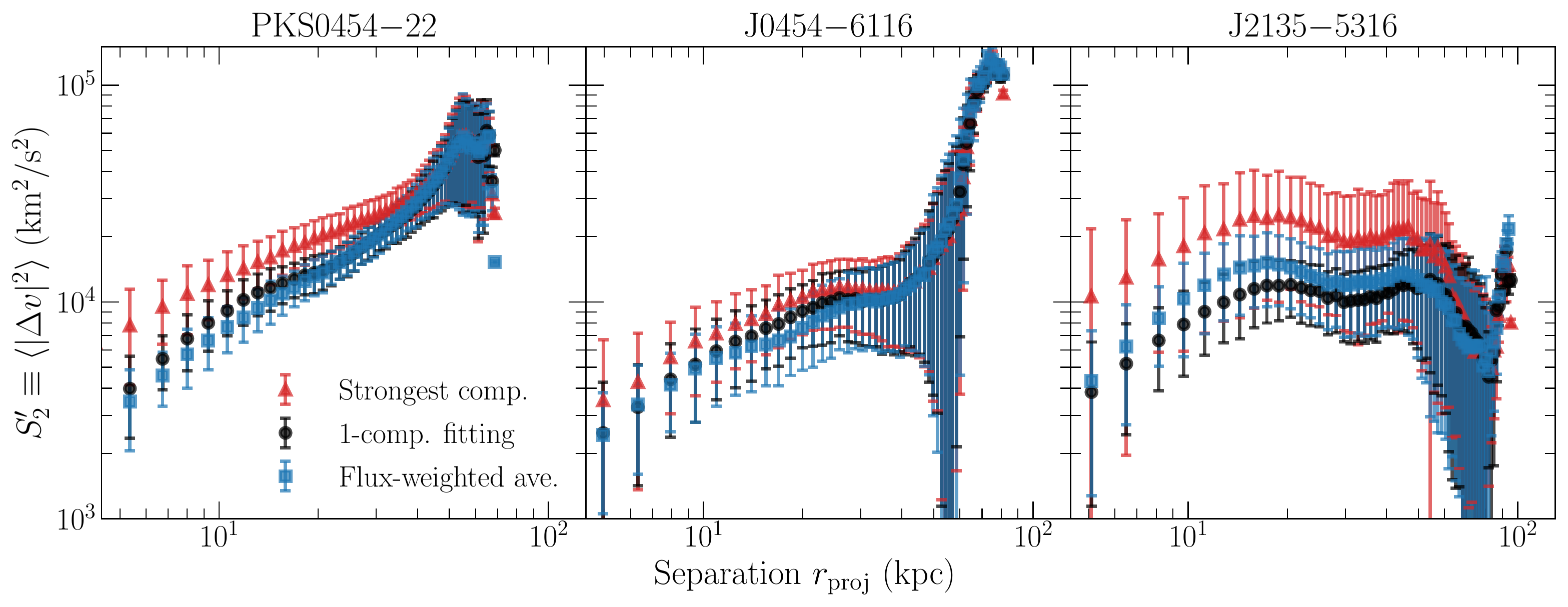}
    \caption{Comparison of the VSF measurement under three different scenarios where we assign to the multi-component spaxels (1) the velocity of the components with the most flux in each spaxel (i.e., the strongest component), (2) the velocity obtained by a one-component fit (i.e., ignoring the presence of multiple components), and (3) the flux-weighted mean velocity among all components. }
    \label{fig:o3_multicomp_compare}
\end{figure*}

\begin{figure*}
    \centering
    \includegraphics[width=\linewidth]{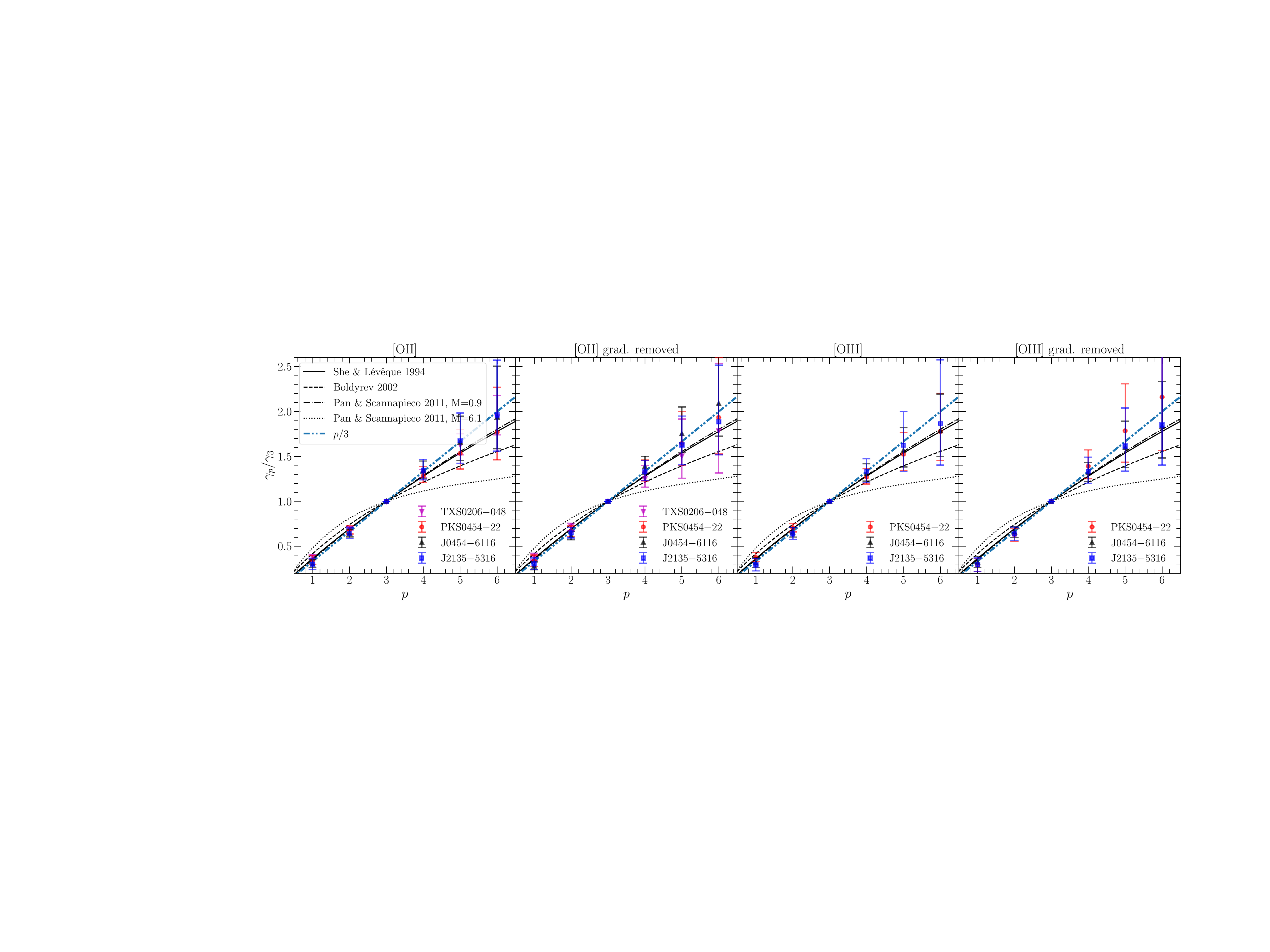}
    \caption{
    Estimated power-law slopes $\gamma_p$ of $S_p$ with $p$ ranging from $p=1$ to $p=6$ for all four nebulae. Different $\gamma_p$'s are normalized to $\gamma_3$ for measurements based on [\ion{O}{II}] $\lambda\lambda\,3727,3729$ and [\ion{O}{III}] $\lambda\,5008$ velocity maps as well as their corresponding velocity residual maps after removing a unidirectional coherent gradient (see \S\,\ref{sec:1Dgrad}). Data points represent the median values obtained with the 1000 modified bootstrap samples (see \S\,\ref{sec:VSF_measurement}), and the error bars indicate the 16$^{\rm th}$ and 84$^{\rm th}$ quantiles. Note that the ratio $\gamma_p / \gamma_3$ only equals to $\gamma_p$ if $\gamma_3=1$. 
    The solid curve shows the expected $\gamma_p / \gamma_3$ ratio for subsonic Kolmogorov turbulence with the intermittency correction presented in \protect\cite{SheLeveque1994}. The expected $\gamma_p / \gamma_3$ ratio for supersonic magnetohydrodynamic turbulence presented in \protect\cite{Boldyrev2002} is shown by the dashed curve.  The dash-dotted (dotted) curve indicates the $\gamma_p / \gamma_3$ 
    ratio derived from numerical hydrodynamic turbulent simulations for Mach number $M=0.9$ ($M=6.1$) as presented in \protect\cite{PanScannapieco2011}.  Finally, the blue loosely dash-dotted curve shows the expected $\gamma_p / \gamma_3$ ratio for Kolmogorov turbulence without the intermittency correction, which simply scales as $p / 3$. See \S\ \ref{sec:ESS_discussion} for further discussions.} 
    \label{fig:ESS_slopes}
\end{figure*}



\bsp	
\label{lastpage}
\end{document}